\documentclass[12pt,draftclsnofoot,onecolumn]{IEEEtran}

\ifCLASSINFOpdf
\else
\fi

\usepackage{graphicx}
\usepackage[cmex10]{amsmath}
\usepackage{hyperref}
\usepackage{amsfonts}
\usepackage{dblfloatfix}
\usepackage{cite}
\usepackage{dsfont}
\usepackage{bm}
\usepackage{amssymb}
\usepackage{array}
\usepackage{algorithm}
\usepackage{algorithmic}
\usepackage{subfigure}
\usepackage{graphics}
\usepackage{epsfig}
\usepackage{epstopdf}
\usepackage{float}
\usepackage{extarrows}
\usepackage{booktabs}
\usepackage{xcolor}

\newtheorem{lemma}{Lemma}

\begin{document}
\abovedisplayskip=1pt
\belowdisplayskip=1pt
\allowdisplaybreaks

\title{Deep-Learned Approximate Message Passing for Asynchronous Massive Connectivity}

\author{
Weifeng~Zhu, Meixia~Tao,~\IEEEmembership{Fellow,~IEEE}, Xiaojun~Yuan,~\IEEEmembership{Senior Member,~IEEE}, and Yunfeng~Guan

\thanks{This paper was presented in part at the IEEE International Conference of Communications (ICC) 2020 \cite{Zhu_2020_ICC}.}
\thanks{W. Zhu, M. Tao, and Y. Guan are with the Department of Electronic Engineering, Shanghai Jiao Tong University, Shanghai 200240,
China (e-mail: \{wf.zhu, mxtao, yfguan69\}@sjtu.edu.cn).}
\thanks{X. Yuan is with the Center for Intelligent Networking and Communication (CINC), University of Electronic Science and Technology of China, Chengdu 610000, China (e-mail: xjyuan@uestc.edu.cn).}
}
\maketitle

\vspace{-1cm}

\begin{abstract}

This paper considers the massive connectivity problem in an asynchronous grant-free random access system, where a huge number of devices sporadically transmit data to a base station (BS) with imperfect synchronization.
The goal is to design algorithms for joint user activity detection, delay detection, and channel estimation.
By exploiting the sparsity on both user activity and delays, we formulate a hierarchical sparse signal recovery problem in both the single-antenna and the multiple-antenna scenarios.
While traditional compressed sensing algorithms can be applied to these problems, they suffer high computational complexity and often require the perfect statistical information of channel and devices.
This paper solves these problems by designing the Learned Approximate Message Passing (LAMP) network, which belongs to model-driven deep learning approaches and ensures efficient performance without tremendous training data.
Particularly, in the multiple-antenna scenario, we design three different LAMP structures, namely, distributed, centralized and hybrid ones, to balance the performance and complexity.
Simulation results demonstrate that the proposed LAMP networks can significantly outperform the conventional AMP method thanks to their ability of parameter learning.
It is also shown that LAMP has robust performance to the maximal delay spread of the asynchronous users.

\end{abstract}
\begin{IEEEkeywords}
Asynchronous massive connectivity, grant-free random access, massive machine-type communication, compressed sensing, approximate message passing, deep learning
\end{IEEEkeywords}

\section{Introduction}\label{sec:introduction}

The fifth generation (5G) of wireless cellular networks has identified the massive machine-type communications (mMTC) as one of its core services \cite{Andrew_2014_JSAC, Bock_2018_Access}. The mMTC service is expected to provide cellular connectivity to a large amount of low-cost machine-type devices for Internet of Things (IoT) applications. A key feature of IoT traffics is that the uplink transmission is usually sporadic and has short packet size, so that only a small and random subset of devices are active and for a short while \cite{Bock_2016_CM, Liu_2018_SPM}. The main challenge to support mMTC services is therefore to design new multiple-access schemes that can facilitate user activity detection, channel estimation and data detection timely and accurately.

Grant-free (GF) random access is promising to establish sporadic connection between machine-type devices and their associated base stations (BSs) with minimal control overhead \cite{Liu_2018_SPM}. In the GF random access, each activated device directly transmits a unique pilot sequence followed by data packets without asking for permission from the BS. The pilot sequences are pre-designed for user identification. They are often non-orthogonal due to a large quantity of devices but limited time-frequency resource.
In each time slot, the BS needs to identify all the active users by detecting which pilots are received, and then estimate their channels for data detection.

Note that many existing works on user activity detection in GF transmission assume that the transmissions of all the active devices are perfectly synchronized \cite{Liu_2018_SPM, Schepker_2013_ISWCS, Wunder_2015_GCW, Chen_2018_TSP, Liu_2018_TSP, Senel_2018_TCOM, Sun_2019_TCOM, Fengler_2020_arxiv, Chen_2019_TWC, Ke_2020_JSAC}.
In practice, low-cost IoT devices usually work in a narrow-band system, and they have bursty transmissions and inconsistent time accuracies. Due to the large overhead, the conventional synchronization mechanism is hard to be employed  among the massive number of low-cost devices. Thus, it is necessary to also consider a scalable scheme to tackle the imperfect synchronization.
If not appropriately handled, such asynchrony in grant-free random access may severely deteriorate the performance of user activity detection and channel estimation.

The goal of this paper is to investigate the joint user activity detection, delay detection and channel estimation in asynchronous massive access systems. By exploiting the sparsity in both the device activity pattern and transmission delay pattern, this paper formulates a hierarchical sparse signal recovery problem.
We solve the problem by designing the Learned Approximate Message Passing (LAMP) network \cite{Borg_2017_TSP} with efficient computation and outstanding recoverability. The LAMP network is obtained by unfolding approximate message passing (AMP) to form a feedforward network, where the parameters in the AMP framework can be learned to enhance the performance with a medium size of training data. Both the single-antenna scenario and the multiple-antenna scenario are considered.



\vspace{-0.4cm}

\subsection{Related Work}

The user activity detection and channel estimation in GF random access schemes are often cast into the sparse signal recovery problems in compressed sensing (CS).
The works \cite{Schepker_2013_ISWCS, Wunder_2015_GCW} propose to jointly detect the active users and estimate their channels without prior knowledge of the channel state information (CSI) by using orthogonal matching pursuit (OMP) and the basis pursuit denoising (BPDN). By exploiting the statistical information of the channels and users, the computationally efficient AMP algorithm \cite{Donoho_2009_PNAS} is adopted in \cite{Chen_2018_TSP, Sun_2019_TCOM, Liu_2018_TSP, Senel_2018_TCOM}.
In particular, the massive MIMO techniques are considered in \cite{Liu_2018_TSP, Senel_2018_TCOM}, and their analysis demonstrates that the user activity detection error can be driven to zero asymptotically when the number of antennas goes to infinity. Recently, a covariance matching algorithm is proposed in \cite{Fengler_2020_arxiv} to be capable of detecting a much larger amount of active users in the massive MIMO scenario.
By exploiting the BS cooperation, the works \cite{Chen_2019_TWC, Ke_2020_JSAC} investigate the sparse activity detection in multi-cell system based on the AMP-based algorithms to further improve the detection performance.
In some cases, however, the BS is only interested in the transmitted information but not the user identification, which thereby motivates the unsourced random access \cite{Poly_2017_ISIT}. In such scenario, all users share a common codebook and the BS applies the CS-based algorithm to decode the transmitted messages without detecting the user activity \cite{Poly_2017_ISIT, Amalladinne_2020_arxiv, Shyianov_2020_JSAC}.
Note that the aforementioned works all assume that the transmit signals from all devices are perfectly synchronized when arriving at the receiver.

Several existing works have attempted to investigate the massive random access problem with imperfect synchronization. In contrast to \cite{Sad_2020_TWC}, the uplink transmissions between different users to the BS are usually considered to be asynchronous in massive access. The work \cite{Abebe_2017_ICC} introduces a blank time interval between the pilot and data whose length is large enough so that the pilot detection and the data detection of different active users will not interfere with each other. In \cite{Ama_2019_ICASSP}, a zero-padding approach is adopted to avoid interference between the data sub-blocks with different indexes from different active users. In both \cite{Abebe_2017_ICC, Ama_2019_ICASSP}, the OMP-based algorithms are proposed to detect the active users, whose complexity is high when the number of users is large. The work \cite{Ding_2019_TWC} introduces a simple signal model with no pilot in the asynchronous systems. Then the user activity and data are jointly detected based on the Turbo bilinear generalized approximate message passing (Turbo-BiG-AMP) algorithm, where the exact statistical information of the system is usually required.

Recently, deep learning (DL) emerges as another powerful approach for user detection and channel estimation by training a deep neural network (DNN) based on a vast amount of labeled training data. Previous works \cite{Bai_2019_VTC, Zhang_2019_TVT} have shown that the DL approaches have potential to offer improvements in both recoverability and complexity.
In contrast to the traditional CS-based algorithm which usually detect the active users based on their estimated channel power, the DL approaches can directly give a more accurate user activity detection solution and then improve the channel estimation further.
In particular, \cite{Bai_2019_VTC} proposes a block-restrict neural network (BRNN) for fast multiuser detection and then performs channel estimation.
On the other hand, the system statistics employed in some iterative algorithms, e.g., message passing-based algorithms, may not be precisely estimated, and the approximation in these algorithms is also likely to be inaccurate in some cases. Therefore, DL can also be integrated in these algorithms to improve the performance by taking advantage of its parameter learning ability.
In the work \cite{Zhang_2019_TVT}, a deep neural network-aided message passing-based block sparse Bayesian learning (DNN-MP-BSBL) algorithm is proposed, which can achieve better channel estimation accuracy with fewer iterations than MP-BSBL.
These existing DL approaches are all designed for synchronous systems. To our best knowledge, DL-based approaches have not be applied to asynchronous systems yet.

\vspace{-0.4cm}

\subsection{Contributions}

This paper considers the joint user activity detection, delay detection and channel estimation in the asynchronous grant-free massive random access system.
We adopt the transmission model that has a guard time inserted between pilot and data signals to capture both the sporadic communication pattern and delay pattern of the asynchronous users, which has been also used before in \cite{Schepker_2013_VTC, Abebe_2017_ICC}.
Then we formulate a hierarchical sparse signal recovery problem based on the signal model. Depending on whether the BS has one or multiple antennas, the joint user activity detection, delay detection and channel estimation can be reformulated as a single measurement vector (SMV) problem or a multiple measurement vector (MMV) problem.
In this work, we propose to solve these problems based on the LAMP network \cite{Borg_2017_TSP} which leverages the deep learning technique and the AMP framework to offer improvements in both recoverability and complexity.
Compared with other DL approaches \cite{Gregor_2010_ICML, Chen_2018_NIPS}, the LAMP network allows feasible performance analysis and achieves better recoverability with more flexible shrinkage function choices. To exploit the common sparsity of channels at all antennas, this paper goes forward to design the LAMP network for the multiple-antenna scenario. Simulation results show that LAMP can outperform AMP by benefitting from the deep learning techniques.
In the multiple-antenna scenario, the LAMP network can be further improved by jointly estimating the channel coefficients on all antennas.
It is also observed that the performance of the proposed LAMP networks is robust to the maximal symbol delay of users.

The main contributions of this work are summarized as follows:
\begin{itemize}
  \item We formulate a hierarchical sparse signal recovery problem with two-level sparsity in the user activity pattern and delay pattern to perform joint user activity detection, delay detection, and channel estimation.
  \item We first design a LAMP network for the single-antenna scenario. Two types of learnable shrinkage functions of the soft thresholding function and the MMSE-optimal denoising function are designed to improve the recovery performance of LAMP by learning their shrinkage parameters from the training data. In particular, the learnable MMSE-optimal denoising function has taken the delay-level sparsity into account, which can further enhance the performance.
      As well, the performance degradation caused by the situation where the measurement matrix has non independent and identically distributed (i.i.d.) elements may also be overcome by the matched filter learning ability of LAMP.
  \item We extend to design the LAMP networks for the multiple-antenna scenario. Three network structures are designed to balance the complexity and recoverability, namely distributed LAMP, centralized LAMP, and hybrid LAMP. Distributed LAMP is designed for the situation with complexity limits, and centralized LAMP achieves the best performance among these three networks and has feasible performance analysis. Hybrid LAMP can balance the performance and complexity to be employed in more complicated systems.
\end{itemize}
\vspace{-0.5cm}

\subsection{Organizations and Notations}

The remaining part of this paper is organized as follows.
Section \ref{sec:system_model} introduces the system model of asynchronous massive connectivity system.
In section \ref{sec:problem_formulation}, we formulate a hierarchical sparse signal recovery problem and introduce the basics of AMP.
In Section \ref{sec:LAMP_SMV}, we introduce the LAMP network design in the single-antenna scenario.
In Section \ref{Sec:LAMP-MMV}, the LAMP networks for the multiple-antenna scenario are designed.
The performance of the proposed approaches is illustrated in Section \ref{Sec:simulation}.
Finally, we conclude this paper in Section \ref{sec:conclusion}.

In this paper, upper-case and lower-case letters denote random variables and their realizations, respectively.
Letters $\mathbf{x}$, $\mathbf{X}$, $\mathcal{X}$ denote vector, matrix and set, respectively.
Superscripts $(\cdot)^T$ denote transpose.
In addition, $\mathbf{1}_N$ denotes the all-one vector with length $N$.
Further, $\mathbb{E}[\cdot]$ denotes the expectation operation; $|\cdot|$ denotes the magnitude of a variable or the Cardinality of a set, depending on the context; $||\cdot||_p$ denotes the $l_p$ norm of a vector; $||\cdot||_F$ denotes the Frobenius norm of a matrix.

\vspace{-0.5cm}
\section{System Model}\label{sec:system_model}

\begin{figure}[t]
  \centering
  \includegraphics[width=.5\textwidth]{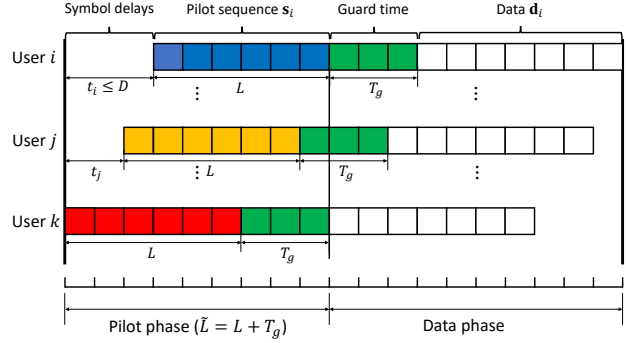}
  \vspace{-0.4cm}
  \caption{Frame structure}\label{fig:frame_model}
  \vspace{-1cm}
\end{figure}

We consider an asynchronous massive access communication system containing one BS with single or multiple antennas and a very large number $N$ users each with single antenna. In each transmission frame, each user sporadically transmits data to the BS with a small and fixed probability $p_a$, and the signals of active users arrive at the BS with different and unknown time delays. The transmit power is identical for each user.
Here, we focus on the system as shown in Fig. \ref{fig:frame_model} where the user signals are asynchronous at the frame level but synchronous at the symbol level. Specifically, the whole frame can be divided into many symbol intervals, and each symbol is exactly at one symbol interval, but the signal sequence may not calibrated at the start of the frame.
Then it means that the delay of each user is an integer number of symbol intervals. The maximal symbol delays of all users are assumed to be much smaller than the frame length.

We adopt a grant-free random access scheme with two-phase transmission including a pilot phase and a data phase. The channel is assumed to be block fading so that it remains unchanged in one frame and varies in different frames. Specifically, the Rayleigh fading channel is considered.
Each user $n$ is assigned with a unique pilot sequence $\mathbf{s}_n = [s_{n,1}, s_{n,2}, \dots, s_{n,L}]^T \in \mathbb{C}^{L \times 1}$ for identification and channel estimation, where $L$ is the length of the pilot sequence and is much smaller than the total number of users $N$, i.e., $L \ll N$. We generate the elements in all pilot sequences from the i.i.d. Gaussian distribution with zero mean and $1/L$ variance, and then normalize each pilot sequence to have unit power.
Denote the symbol delay of the signal of user $n$ by $t_n$ when it arrives at the BS. We assume that $t_n$ is unknown, discrete, and uniformly distributed in the set $\{0,1,\dots,D \}$ with $D$ being the maximal symbol delay spread. Note that there is no power transmission in the guard interval. The frame structure containing a guard time between the pilot and data as shown in Fig. \ref{fig:frame_model} is introduced in this work, where the length of the guard time, denoted as $T_g \in \mathbb{Z}$, is chosen to be equal to or larger than $D$, i.e., $T_g \ge D$. Based on such $T_g$, the pilot transmission of each user will not overlap the data transmission of all other users. Though the maximal delay may not be obtained accurately, we can set $T_g$ a sufficiently large value. In this paper, we simply set the length of the guard time to be equal to the known maximal delay, i.e., $T_g=D$, without affecting the algorithm design.
Here, we assume that the BS does not know the explicit value of the user active probability and the channel informationm, while we have pre-collected a large data set and these system statistics can be implicitly exploited.
Let $\lambda_n \in \{1,0\}$ denote whether user $n$ is active or not.
We take both the random user activity $\lambda_n$ and the unknown symbol delays $t_n$ into account in the signal model.
For notation simplicity in the following, we define $\tilde{L} = L+T_g$ and $\tilde{N} = N(T_g+1)$, where $\tilde{L}$ can be viewed as the expanded pilot length.

When there is one single antenna equipped at the BS, the received signal at the BS in the pilot phase can be expressed as
\begin{equation}\label{equ:rece_sig_SMV}
    \mathbf{y} = \sum_{n=1}^{N} \tilde{\mathbf{s}}_{n,t_n+1} \lambda_n h_n + \mathbf{z},
\end{equation}
where $\tilde{\mathbf{s}}_{n,t_n+1} \in \mathbb{R}^{\tilde{L} \times 1}$ is the expanded pilot sequence of user $n$ obtained by adding $t_n$ zeros before $\mathbf{s}_n$ and $(T_g-t_n)$ zeros after $\mathbf{s}_n$, i.e., $\tilde{\mathbf{s}}_{n,t_n+1} = [\mathbf{0}^T_{t_n}, \mathbf{s}_n^T, \mathbf{0}^T_{T_g-t_n}]^T$; $h_n = \sqrt{\phi_n} g_n \in \mathbb{R}$ denotes the channel coefficient between the BS and user $n$ where $\phi_n$ represents the large-scale fading attenuation and $g_n$ is the normalized small-scale fading with zero mean; $\mathbf{z} \in \mathbb{R}^{\tilde{L} \times 1}$ denotes the additive Gaussian noise vector with variance $\sigma^2_z$ normalized by the transmit power.

Define $\tilde{\pmb\lambda}_n = [\lambda_{n,1}, \lambda_{n,2}, \dots, \lambda_{n,T_g+1}]^T \in \mathbb{R}^{(T_g+1) \times 1}$ to indicate both the activity state and symbol delay of user $n$, where there is at most one non-zero entry. In specific, we has $\lambda_{n,t}=0, \forall t \in \{1,2,...,T_g+1\}$ if user $n$ is inactive, otherwise if user $n$ is active and has a symbol delay of $t_n$, we have
\begin{equation}\label{equ:ind_au}
    \lambda_{n,t} = \left\{ \begin{array}{ll}
                              1, & t=t_n+1,  \\
                              0, & \textrm{otherwise}.
                            \end{array} \right.
\end{equation}

The received signal $\mathbf{y}$ in (\ref{equ:rece_sig_SMV}) can be rewritten in a matrix-vector form as
\begin{equation}\label{equ:rece_sig_mat_SMV}
    \mathbf{y} = \tilde{\mathbf{S}} \tilde{\mathbf{\Lambda}} \mathbf{h} + \mathbf{z} = \tilde{\mathbf{S}} \mathbf{x} + \mathbf{z},
\end{equation}
where $\tilde{\mathbf{S}} = [\tilde{\mathbf{S}}_1, \tilde{\mathbf{S}}_2, \dots, \tilde{\mathbf{S}}_N] \in \mathbb{R}^{\tilde{L} \times \tilde{N}}$ refers to the expanded pilot matrix of all users in the system with $\tilde{\mathbf{S}}_n = [\tilde{\mathbf{s}}_{n,1}, \tilde{\mathbf{s}}_{n,2}, \dots, \tilde{\mathbf{s}}_{n,T_g+1}] \in \mathbb{R}^{\tilde{L} \times (T_g+1)}$ denoting the expanded pilot matrix of user $n$;
$\tilde{\mathbf{\Lambda}} = \mathbf{diag}(\tilde{\mathbf{\Lambda}}_1, \tilde{\mathbf{\Lambda}}_2, \dots, \tilde{\mathbf{\Lambda}}_N) \in \mathbb{R}^{\tilde{N} \times \tilde{N}}$ is the indicator matrix with  $\tilde{\mathbf{\Lambda}}_n = \mathbf{diag}(\tilde{\mathbf{\pmb\lambda}}_n) \in \mathbb{R}^{(T_g+1) \times (T_g+1)}$;
$\mathbf{h} = [h_1 \mathbf{1}_{T_g+1}^T, h_2 \mathbf{1}_{T_g+1}^T, \dots, h_N \mathbf{1}_{T_g+1}^T]^T \in \mathbb{R}^{\tilde{N} \times 1}$ is the overall channel vector; finally, $\mathbf{x} = \tilde{\mathbf{\Lambda}} \mathbf{h} \in \mathbb{R}^{\tilde{N} \times 1}$ represents the effective channel vector of all users that contains the information of user activity pattern, the channel gains and the symbol delays.

Likewise, when there are $M$ antennas at the BS, the received signal $\mathbf{Y}$ during the pilot phase can be written as
\begin{equation}\label{equ:rece_sig_mat_MMV}
\mathbf{Y} = \tilde{\mathbf{S}} \tilde{\mathbf{\Lambda}} \mathbf{H} + \mathbf{Z} = \tilde{\mathbf{S}} \mathbf{X} + \mathbf{Z},
\end{equation}
where $\mathbf{H} = [\mathbf{h}_1 \mathbf{1}_{T_g+1}^T, \mathbf{h}_2 \mathbf{1}_{T_g+1}^T, \dots, \mathbf{h}_N \mathbf{1}_{T_g+1}^T]^T \in \mathbb{R}^{\tilde{N} \times M}$ is the overall channel matrix and $\mathbf{h}_n = \sqrt{\phi_n}\mathbf{g}_n \in \mathbb{R}^{M \times 1}$ is the channel vector between the BS and user $n$ with $\mathbf{g}_n$ denoting the small-scale fading vector, and each element in $\mathbf{g}_n$ is assumed to be i.i.d.; $\mathbf{X} = \tilde{\mathbf{\Lambda}} \mathbf{H}  \in \mathbb{R}^{\tilde{N} \times M}$ represents the effective channel matrix of all users; finally, $\mathbf{Z} \in \mathbb{R}^{\tilde{L} \times M}$ denotes the additive Gaussian noise matrix normalized by the transmit power.

\begin{figure}[t]
  \centering
  \includegraphics[width=.5\textwidth]{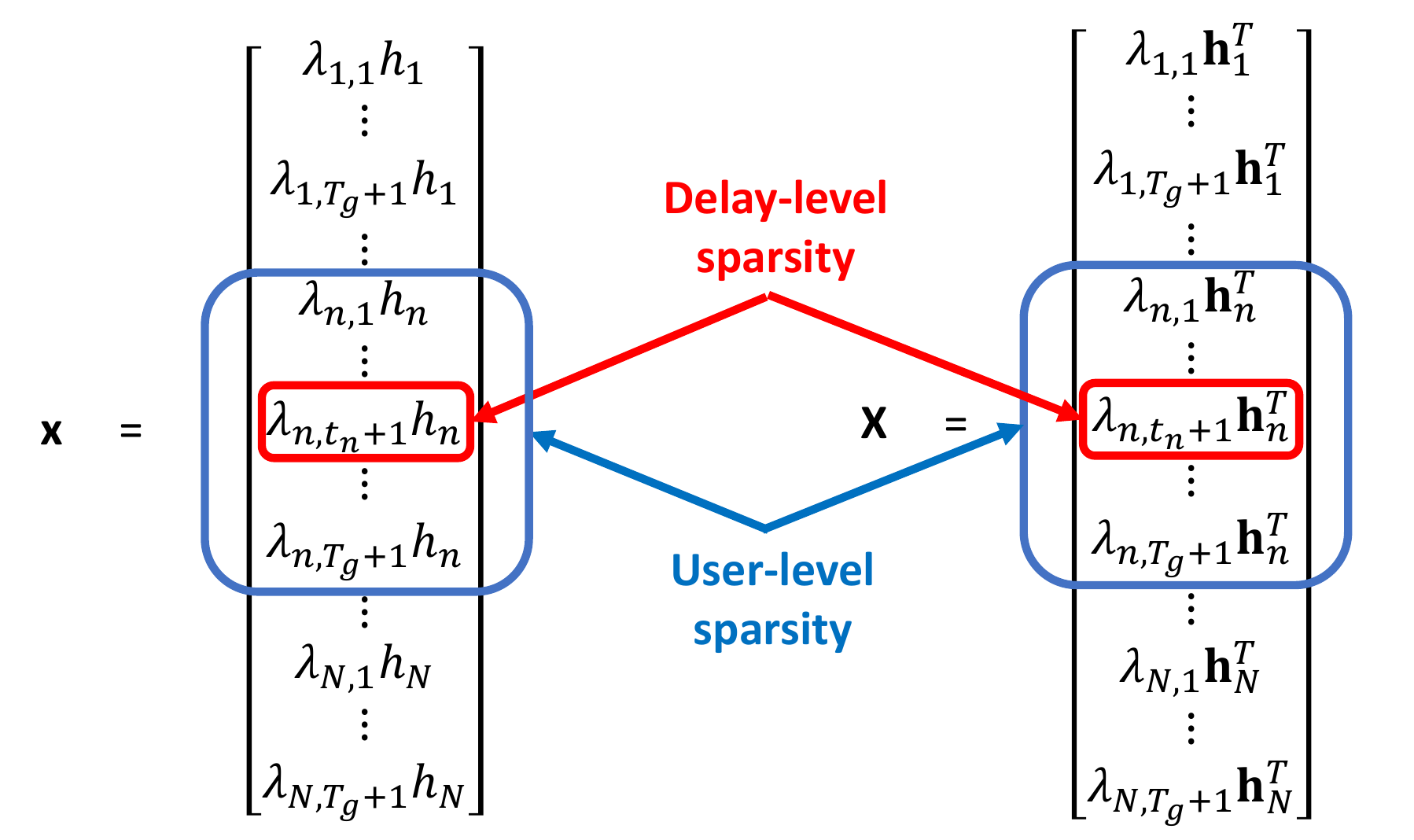}
  \vspace{-0.4cm}
  \caption{Two-level sparse structure of the effective channel.}\label{fig:X_model}
  \vspace{-1cm}
\end{figure}

The problem is to recover $\mathbf{x}$ or $\mathbf{X}$ from the received signal $\mathbf{y}$ in (\ref{equ:rece_sig_mat_SMV}) or $\mathbf{Y}$ in (\ref{equ:rece_sig_mat_MMV}), respectively, given the measurement matrix $\tilde{\mathbf{S}}$. This is a classic underdetermined linear inverse problem in the SMV form or the MMV form. Due to the sporadic communication pattern and only one single symbol delay of each asynchronous user, the effective channel $\mathbf{x}$ or $\mathbf{X}$ has a hierarchical sparse representation. As shown in Fig. \ref{fig:X_model}, the hierarchical sparsity consists of two levels of sparsity including the \emph{user-level sparsity} and the \emph{delay-level sparsity}.
The user-level sparsity means that most $\mathbf{x}_{n} = [x_{n,1}, \dots, x_{n,T_g+1}]^T = [\lambda_{n,1}h_n, \dots, \lambda_{n,T_g+1}h_n]^T$ in (\ref{equ:rece_sig_mat_SMV}) or $\mathbf{X}_{n} = [\mathbf{x}_{n,1}, \dots, \mathbf{x}_{n,T_g+1}]^T =  [\lambda_{n,1}\mathbf{h}_n, \dots, \lambda_{n,T_g+1}\mathbf{h}_n]^T$ in (\ref{equ:rece_sig_mat_MMV}) are zeros, and the delay-level sparsity enforces that there is only one non-zero element in the non-zero $\mathbf{x}_n$ or only one non-zero row in the non-zero $\mathbf{X}_n$.
Therefore, the underdetermined problem can be possibly solved based on the CS algorithms.\footnote{Though the measurement matrix $\tilde{\mathbf{S}}$ may not satisfy the restricted isometry property (RIP), simulation results show that the problem can be well solved by CS-based algorithms.}

\vspace{-0.5cm}
\section{Problem Formulation and AMP Algorithm}\label{sec:problem_formulation}

In this section, we formulate our problem of joint user activity detection, delay detection and channel estimation in both the SMV and MMV forms. Then we introduce the AMP algorithm to solve the formulated problems.

\vspace{-0.4cm}
\subsection{SMV problem for single antenna}

We first consider the problem in the single-antenna scenario, which is an SMV problem.
By considering the user-level sparsity and the delay-level sparsity in the effective channel vector, the problem is formulated as
\begin{subequations}\label{opt:original_prob_SMV}
	\begin{align}
	\min_{\mathbf{x}}\quad & ||\mathbf{y} - \tilde{\mathbf{S}} \mathbf{x}||_2^2 \\
	\text{s.t.}\quad & ||\mathbf{x}||_0 \leq C, \label{opt:group_sparsity} \\
	\quad       & ||\mathbf{x}_n||_0 \leq 1, n = 1, 2, \dots, N, \label{opt:element_sparsity}
	\end{align}
\end{subequations}
where the constant $C \ll N$ denotes the maximal number of active users in one frame. The constraint (\ref{opt:group_sparsity}) comes from the user-level sparsity with $||\mathbf{x}||_0 = \sum_{n=1}^{N} ||\mathbf{x}_n||_0$ and $||\mathbf{x}_n||_0$ indicates whether user $n$ is active or not. The constraint (\ref{opt:element_sparsity}) implies the delay-level sparsity that ensures that each active user has only one single symbol delay.

The problem (\ref{opt:original_prob_SMV}) is very difficult to solve due to the non-smooth constraints. Then we reformulate the problem as
\begin{equation}\label{opt:original_prob_lag_SMV}
\min_{\mathbf{x}} \beta ||\mathbf{x}||_0 + \frac{1}{2} ||\mathbf{y} - \tilde{\mathbf{S}} \mathbf{x}||_2^2.
\end{equation}
Note that the problem (\ref{opt:original_prob_lag_SMV}) can give a sparse channels $\mathbf{x}$ close to the optimal solution of (\ref{opt:original_prob_SMV}), though the constraint (\ref{opt:element_sparsity}) cannot surely guaranteed. To refine the solution of (\ref{opt:original_prob_lag_SMV}), we introduce an common element selection operation to enforce all the elements except the one with the largest magnitude in each group to be zeros after obtaining the solutions to the problem. Then user activity detection can be performed by comparing the channel power with a predefined decision threshold $q_{th}$. The trade-off between the sparsity of the solution and the mean square error (MSE) $||\mathbf{y}-\tilde{\mathbf{S}}\mathbf{x}||^2_2$ is adjusted by the value of the tuning parameter $\beta$.
We then adopt the common strategy to relax the objective function into a convex function by using the $l_1$ norm as
\begin{equation}\label{opt:relaxed_prob_SMV}
\min_{\mathbf{x}}\quad \beta ||\mathbf{x}||_1 + \frac{1}{2} ||\mathbf{y} - \tilde{\mathbf{S}} \mathbf{x}||_2^2.
\end{equation}

The problem (\ref{opt:relaxed_prob_SMV}) is a LASSO problem and can be directly solved by BPDN using interior point method with computation complexity $\mathcal{O}(\tilde{N}^{3.5})$.
The OMP-based greedy methods can be used as well, but they involve the matrix inverse operation in each iteration. The computation complexity is $\mathcal{O}(K^3)$ in each iteration with $K$ being the expected number of active users, which is still high in a large-scale system. The recently proposed AMP-based algorithms have much lower complexity in the large-scale system while approaches the performance of LASSO asymptotically, given by $\mathcal{O}(\tilde{L}\tilde{N})$ in each iterations.
Thus, we propose to employ the AMP-based algorithms in our problem.

The AMP algorithm for our considered problem in the single-antenna scenario is described as follows. It starts with $\widehat{\mathbf{x}}_0 = \mathbf{0}$ and $\mathbf{v}_{0} = \mathbf{y}$ and performs the following computations in the $i$th iteration \cite{Donoho_2009_PNAS}
\begin{subequations}\label{Alg:AMP-SMV_iteration}
	\begin{align}
    \widehat{\mathbf{r}}_{i} & = \widehat{\mathbf{x}}_{i-1} + \tilde{\mathbf{S}}^T \mathbf{v}_{i-1}, \label{equ:r_est} \\
	\widehat{\mathbf{x}}_{i} & = \boldsymbol{\eta}(\widehat{\mathbf{r}}_{i}; \sigma_{i-1}, \pmb\vartheta), \label{equ:x_estimate} \\
    \mathbf{v}_{i}                    & = \mathbf{y} - \tilde{\mathbf{S}} \widehat{\mathbf{x}}_{i} + \frac{1}{\tilde{L}} b_{i} \mathbf{v}_{i-1}, \label{equ:residual}
	\end{align}
\end{subequations}
where $\sigma_{i-1} = \frac{1}{\sqrt{\tilde{L}}} ||\mathbf{v}_{i-1}||_2$ is the estimated standard deviation of the corrupted noise in $\widehat{\mathbf{r}}_{i}$; the variable $b_{i} = \sum_{n=1}^{N}\sum_{t=1}^{T_g+1}\frac{\partial[\boldsymbol\eta(\mathbf{r}; \sigma_{i-1}, \pmb\vartheta)]_{n,t}}{\partial r_{n,t}} \Big|_{\mathbf{r} = \widehat{\mathbf{r}}_{i}}$ is calculated to obtain ``Onsager correction'' term $\frac{1}{\tilde{L}} b_{i} \mathbf{v}_{i-1}$;
and the shrinkage function $\boldsymbol{\eta}(\cdot)$ is usually a non-linear component-wise function operating on each element of $\widehat{\mathbf{r}}_i$ individually with $\pmb\vartheta$ being the parameter set. 
Note that any Lipschitz-continuous shrinkage function can be used \cite{Borg_2017_TSP}, which enables the usage of more well-performed shrinkage functions if we have more prior knowledge of the channels and users. For example, the shrinkage function based on MMSE-optimal criterion can be designed to achieve better recoverability when the system statistics are perfectly known. Since there exists the delay-level sparsity in the effective channel vector $\mathbf{x}_n$, which means that the common component-wise function may not be able to achieve the optimal performance of AMP. In the following, we have designed a non-separable shrinkage function based on the MMSE-optimal criterion to also exploit the delay-level sparsity in the effective channel $\mathbf{x}$.
In AMP, the ``Onsager correction'' term enables the algorithm to be analyzable by state evolution in the asymptotic regime, i.e., $\tilde{L}$, $\tilde{N} \to \infty$ with their ratio and $p_a$ fixed when $\tilde{\mathbf{S}}$ has i.i.d. sub-Gaussian elements \cite{Donoho_2009_PNAS}. The state evolution is given by
\begin{equation}\label{equ:st_SMV}
    \delta^2_{i} = \frac{1}{\tilde{L}}\mathbb{E}_{\mathbf{d}}[||\pmb\eta_{\pmb\vartheta,\delta_{i-1}}(\mathbf{x}^0+\delta_{i-1}\mathbf{d}) - \mathbf{x}^0||_2^2]+\sigma_z^2,
\end{equation}
where $\mathbf{d}$ are the random variables with distribution $\mathcal{N}(0,\mathbf{I}_{\tilde{N}})$. The input to the shrinkage function in the $i$th iteration can be modeled as an AWGN-corrupted version of the true vector $\mathbf{x}^0$ plus noise with estimated variance $\sigma^2_{i-1}$, i.e., $\widehat{\mathbf{r}}_i = \mathbf{x}^0 + \delta_{i-1}\mathbf{d}$.
The variable $\delta_{i-1}$ is usually estimated by the empirical result $\sigma_{i-1}$, i.e., $\widehat{\delta}_{i-1}=\sigma_{i-1} $.

\vspace{-0.5cm}
\subsection{MMV problem for multiple antennas}\label{subSec:MMV_problem}

When the BS has multiple antennas, the considered problem becomes an MMV problem, which is formulated as
\begin{subequations}\label{opt:original_prob_MMV}
    \begin{align}
      \min_{\mathbf{X}}\quad & ||\mathbf{Y} - \tilde{\mathbf{S}} \mathbf{X}||_F^2 \\
      \text{s.t.}  \quad & \sum_{n=1}^{N} \sum_{t=1}^{T_g+1} \mathbb{I} (\mathbf{x}_{n,t}) \leq C, \label{opt:block_sparsity_re} \\
      \quad        & \sum_{t=1}^{T_g+1}\mathbb{I} (\mathbf{x}_{n,t}) \leq 1, n = 1, 2, \dots, N. \label{opt:row_sparsity_re}
    \end{align}
\end{subequations}
where $\mathbb{I}(\cdot)$ is the indicator function with boolean output defined as
\begin{equation}\label{equ:ind_func}
    \mathbb{I}(\mathbf{x}) = \left\{ \begin{array}{ll}
                              1, & \textrm{if $\mathbf{x}$ has non-zero elements},  \\
                              0, & \textrm{otherwise}.
                            \end{array} \right.
\end{equation}
The problem (\ref{opt:original_prob_MMV}) is also difficult to solve directly. By following the similar operations in the SMV problem, we can reformulate the problem as
\begin{equation}\label{opt:relaxed_problem_MMV}
\min_{\mathbf{X}} ~ \beta  \sum_{n=1}^{N} \sum_{t=1}^{T_g+1} ||\mathbf{x}_{n,t}||_2 + \frac{1}{2} ||\mathbf{Y} - \tilde{\mathbf{S}} \mathbf{X}||_F^2.
\end{equation}

We replace the indicator function using the $l_2$ norm and the problem is relaxed as a $l_{2,1}$-norm RLS problem \cite{Tropp_2006_SP}. To solve the relaxed problem, the conventional convex problem solver and greedy algorithm still suffer high computation complexity. Two recently proposed AMP-based algorithms can be applied to solve the MMV problem  (\ref{opt:relaxed_problem_MMV}) with affordable complexity.

\subsubsection{Parallel AMP-MMV}

The parallel AMP-MMV algorithm proposed in \cite{Ziniel_2013_TSP} solves the MMV problem in the distributed way. In each iteration, the parallel AMP-MMV algorithm first estimates the channel coefficients on the $M$ antennas separately and then exchanges the soft information of user activity among different antennas. In specific, each iteration in parallel AMP consists of four distinct phases, which can be labeled using the mnemonics (into), (within), (out), and (across). In phase (into), the current beliefs about the user activity are calculated and conveyed into each AMP-SMV solver. In phase (within), Each of the $M$ AMP-SMV solvers with the current beliefs follows (\ref{Alg:AMP-SMV_iteration}) to solve the SMV problem of estimating the effective channel coefficients on the corresponding antenna in parallel. In phase (out), the solution is utilized to refine the beliefs of the user activity in each AMP-SMV solver. Finally, in phase (across), the beliefs are conveyed across different AMP-SMV solvers. The algorithm will terminate after several iterations. Interested readers may refer to \cite{Ziniel_2013_TSP, Chen_2018_TSP} for more details.

\subsubsection{AMP with vector shrinkage function}

The AMP algorithm equipped with the vector shrinkage functions proposed in \cite{Kim_2011_arxiv} solves the MMV problem in the centralized way. Similar to the iterative procedure in (\ref{Alg:AMP-SMV_iteration}), this algorithm starts with $\widehat{\mathbf{X}}_0 = \mathbf{0}$ and $\mathbf{V}_{0} = \mathbf{Y}$, then computes in the $i$th iteration
\begin{subequations}\label{Alg:AMP-MMV_iteration}
	\begin{align}
    \widehat{\mathbf{R}}_{i} & = \widehat{\mathbf{X}}_{i-1} + \tilde{\mathbf{S}}^T \mathbf{V}_{i-1}, \label{equ:R_est} \\
    \widehat{\mathbf{X}}_{i} & = \boldsymbol{\eta}(\widehat{\mathbf{R}}_{i}; \sigma_{i-1}, \pmb\vartheta), \label{equ:X_est} \\
    \mathbf{V}_{i}                 & = \mathbf{Y} - \tilde{\mathbf{S}} \widehat{\mathbf{X}}_{i} + \frac{1}{\tilde{L}} \mathbf{V}_{i-1} \mathbf{B}_{i}, \label{equ:Residual}
	\end{align}
\end{subequations}
where $\mathbf{R} = [\mathbf{R}_{1}^T, \dots, \mathbf{R}_{N}^T]^T$ with $\mathbf{R}_n = [\mathbf{r}_{n,1}^T, \dots, \mathbf{r}_{n,T_g+1}^T]^T$ is the input to the vector shrinkage function; $\sigma_{i-1} = \frac{1}{\sqrt{\tilde{L}M}} ||\mathbf{V}_{i-1}||_F$ is the estimated standard deviation of the corrupted noise; and $\mathbf{B}_{i} = \sum_{n=1}^{N}\sum_{t=1}^{T_g+1}\frac{\partial[\boldsymbol\eta(\mathbf{R}; \sigma_{i-1}, \pmb\vartheta)]_{n,t}}{\partial \mathbf{r}_{n,t}} \Big|_{\mathbf{R} = \widehat{\mathbf{R}}_{i}}$ is calculated for the ``Onsager correction'' matrix $\frac{1}{\tilde{L}} \mathbf{V}_{i-1} \mathbf{B}_{i}$.
This algorithm in the asymptotic region can also be analyzed by the state evolution expressed as $\pmb\Sigma_{i} = \delta^2_{i}\mathbf{I}$ with $\delta^2_{i}$ determined by
\begin{equation}\label{equ:se_MMV}
    \delta^2_{i} = \frac{1}{\tilde{L}M}\mathbb{E}_{\mathbf{D}}[||\pmb\eta_{\pmb\vartheta,\delta_{i-1}}(\mathbf{X}^0+\delta_{i-1}\mathbf{D}) - \mathbf{X}^0||_F^2]+\sigma_z^2,
\end{equation}
where each element in the random matrix $\mathbf{D} \in \mathbb{R}^{\tilde{N} \times M}$ satisfies i.i.d. Gaussian distribution $\mathcal{N}(0,1)$. The input to the vector shrinkage function in the $i$th iteration can also be modeled as an AWGN-corrupted signal, i.e., $\widehat{\mathbf{R}}_i = \mathbf{X}^0 + \delta_{i-1}\mathbf{D}$. As well, $\delta_{i-1}$ can be estimated by $\sigma_{i-1}$.

\section{Deep-Learned AMP in the Single-Antenna Scenario}\label{sec:LAMP_SMV}

Deep learning is also a powerful approach to accurately estimate the sparse vector $\mathbf{x}$ from the received signal $\mathbf{y}$, where the network parameters of the DNN are trained to minimize the reconstruction MSE by utilizing a large amount of training data $\{(\mathbf{y}^d_T,\mathbf{x}^d_T)\}_{d=1}^{D_T}$ regarded as (feature, label) pairs. Once the training process is completed, the DNN can predict the unknown channels $\mathbf{x}_{New}$ with the newly received signal $\mathbf{y}_{New}$.
In this section, we introduce the LAMP network that combines the DL techniques with the AMP framework to solve our problem in the single-antenna scenario.


\begin{figure}[t]
  \centering
  \includegraphics[width=.5\textwidth]{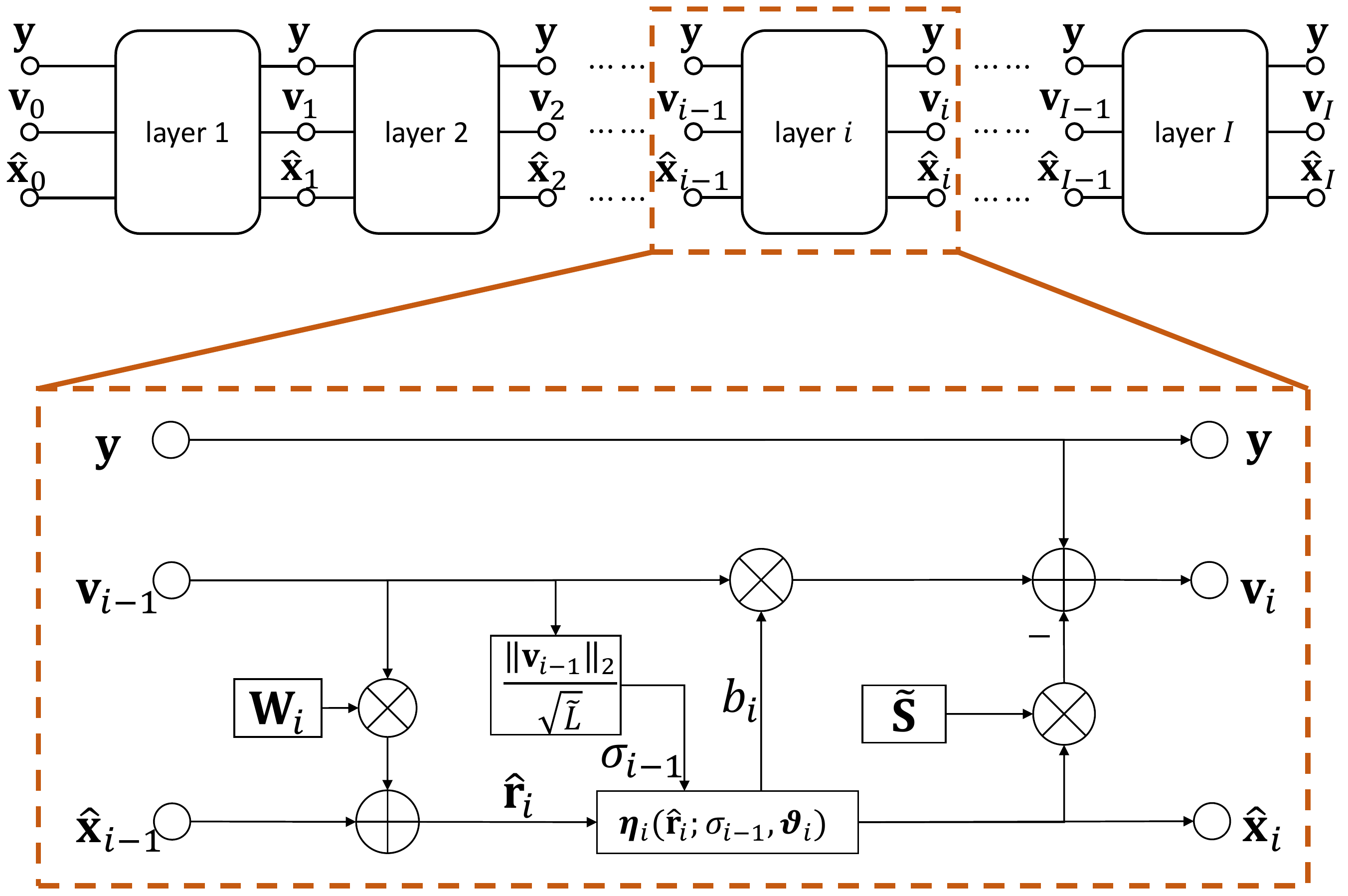}
  \vspace{-0.3cm}
  \caption{The network structure and the $i$th layer of the LAMP network in the single-antenna scenario}\label{fig:LAMP_net}
  \vspace{-0.8cm}
\end{figure}

\subsection{Network structure}

The LAMP network is built by unfolding the iterations of AMP from (\ref{Alg:AMP-SMV_iteration}) as a feedforward neural network. The structure of the LAMP network and the details of the $i$th layer are shown in Fig. \ref{fig:LAMP_net}, which is different from the traditional deep neural networks containing multiple layers of perceptions.
The signal flow graph of the LAMP network is the same as that of the AMP algorithm, where $\mathbf{W}_i$ can be regarded as the matched filter matrix and $\pmb\vartheta_i$ is the shrinkage parameter set. In the LAMP network, $\{\mathbf{W}_i, \pmb{\vartheta}_i\}_{i=1}^{I}$ are considered as the network parameters, which will be learned from the training data.
In the training process, the MSE loss function, defined as $\mathcal{L}(\widehat{\mathbf{x}}) = ||\widehat{\mathbf{x}}-\mathbf{x}^0||_2^2$, is used and the values of the learnable parameters will be updated by following the back-propagation rule \cite{Rume_1986_PDP}. The back-propagation rule calculates the gradients $\frac{\partial \mathcal{L}(\mathbf{W}_i,\pmb{\vartheta}_i)}{\partial\mathbf{W}_i}$ and $\frac{\partial \mathcal{L}(\mathbf{W}_i,\pmb{\vartheta}_i)}{\partial\pmb{\vartheta}_i}$ under the chain rule with the input data batch, and then updates the learnable parameters based on the gradient descending methods.


In LAMP, the shrinkage function also plays an important role. Apart from the commonly used soft thresholding shrinkage (ST) function and the MMSE-optimal denoising function, many other functions, such as the piecewise linear function and the spline function, can be employed as well in LAMP. It is mentioned that the learnable shrinkage parameters in different layers of LAMP can have different values, which can improve the denoising ability of each layer. In this paper, both the ST function and the MMSE-optimal denoising function are considered.
We omit the iteration index $i$ in the following for simplicity. The ST function is given by
\begin{equation}\label{fun:st_SMV}
    [\pmb{\eta}(\widehat{\mathbf{r}})]_{n,t} = \Big( \widehat{r}_{n,t} - \theta_{n,t}\sigma \frac{\widehat{r}_{n,t}}{|\widehat{r}_{n,t}|} \Big) \cdot \mathbb{I}(|\widehat{r}_{n,t}| >\theta_{n,t} \sigma),
\end{equation}
where $\theta_{n,t}$ is a learnable tuning parameter and $\sigma$ is the estimated variance of the corrupted noise in $\widehat{\mathbf{r}}$. The learnable shrinkage parameter set is obtained as $\pmb\vartheta = \{ \theta_{n,t} \}_{n=1,t=1}^{N,T_g+1}$. Note that in AMP, the value of $\theta_{n,t}$ is usually obtained empirically and hence may not be optimal, while the optimal value of $\theta_{n,t}$ can be learned in LAMP. Usually, all the tuning parameters share the same value, i.e., $\theta_{n,t}=\theta, \forall n,t$. Since the symbol delay of each user is assumed to be uniformly distributed in $\{0,1,\dots,T_g\}$, the active probability for each user with delay $t_n$ is obtained as $p_{a,T} = \frac{p_a}{T_g+1}$. Then the effective channel $\mathbf{x}_n$ of user $n$ can be modeled as a joint distribution as
\begin{equation}\label{equ:x_BG_SMV}
    p(\mathbf{x}_{n}) = (1-p_{a})\prod_{t=1}^{T_g+1}\delta(x_{n,t}) + p_{a,T}\sum_{t=1}^{T_g+1}\left(\mathcal{CN}(x_{n,t};0,\phi_n)\prod_{t'\ne t}\delta(x_{n,t'})\right).
\end{equation}
So that the MMSE-optimal denoising function is designed to recover all elements $x_{n,t}$ in each $\mathbf{x}_n$ altogether but not treats them independently as the ST function does. In this way, the delay-level sparsity has been considered in the LAMP network. Based on the joint probability (\ref{equ:x_BG_SMV}), the MMSE-optimal denoising function is given as
\begin{align}\label{equ:shrinkage_MMSE}
    \widehat{x}_{n,t} &= \mathbb{E}[x_{n,t}|\widehat{\mathbf{r}}_{n}] \notag \\
    \quad &= \frac{\frac{\mathcal{CN}(\widehat{r}_{n,t};0,\phi_n+\sigma^2)}{\mathcal{CN}(\widehat{r}_{n,t};0,\sigma^2)} \widehat{r}_{n,t}} {(1+\frac{\sigma^2}{\phi_n}) \left(\sum_{t'=1}^{T_g+1} \frac{\mathcal{CN}(\widehat{r}_{n,t'};0,\sigma^2+\phi_n)}{\mathcal{CN}(\widehat{r}_{n,t'};0,\sigma^2)}
  + \frac{1-p_a}{p_{a,T}}\right)} \notag \\
    \quad &= \frac{\exp\bigg(\frac{|\widehat{r}_{n,t}|^2}{\sigma^2(1+\frac{\sigma^2}{\phi_n})}\bigg) \widehat{r}_{n,t} }{(1+\frac{\sigma^2}{\theta_{n}}) \left[\exp\bigg(\frac{||\widehat{\mathbf{r}}_{n}||_2^2}{\sigma^2(1+\frac{\sigma^2}{\phi_n})}\bigg) + (1+\frac{\phi_n}{\sigma^2})\frac{1-p_a}{p_{a,T}}\right]}.
\end{align}

After obtain equation (\ref{equ:shrinkage_MMSE}), the learnable MMSE-optimal denoising function can then be defined as
\begin{equation}\label{equ:shrinkage_MMSE_learn}
    [\pmb{\eta}(\widehat{\mathbf{r}})]_{n,t} = \theta_3 \frac{\exp\bigg(\frac{|\widehat{r}_{n,t}|^2}{\sigma^2(1+\frac{\sigma^2}{\theta_{1,n}})}\bigg) \widehat{r}_{n,t} }{(1+\frac{\sigma^2}{\theta_{n}}) \left[\exp\bigg(\frac{||\widehat{\mathbf{r}}_{n}||_2^2}{\sigma^2(1+\frac{\sigma^2}{\theta_{1,n}})}\bigg) + \big(1+\frac{\theta_{1,n}}{\sigma^2}\big)\theta_{2}\right]} - \theta_4 \widehat{r}_{n,t},
\end{equation}
where the learnable shrinkage parameter set is defined as $\pmb\vartheta = \{ \{\theta_{1,n}\}_{n=1}^{N}, \theta_{2}, \theta_{3}, \theta_{4}\}$ and we set $\theta_{1,n} = \phi_{n}$ and $\theta_{2}=\frac{1-p_{a}}{p_{a,T}}$. The parameters $\theta_3$ and $\theta_4$ are regarded as the tuning parameters to mix the linear and non-linear shrinkage functions, which is promising to improve the performance of (\ref{equ:shrinkage_MMSE}).
When applying the MMSE-optimal denoising function in the traditional AMP algorithm, we usually set $\theta_3 = 1, \theta_4 = 0$ and determine the value of $\{\{\theta_{1,n}\}_{n=1}^{N}, \theta_2\}$ based on the perfectly known system statistical parameters $\{\{\phi_{n}\}_{n=1}^{N}, p_{a}\}$. In the LAMP network, these parameters are all learned from the training data.
If all users have the same large-scale channel attenuation, we can set $\theta_{1,n} = \theta_1, \forall n=1,\dots,N$. Thus, the number of the learnable parameters can be significantly reduced, given that $N$, the total number of users, is very large.
The calculation of the Onsager term in the neural network needs the derivative of the shrinkage function, which is given by
\begin{equation}\label{equ:derivative_sca_SMV_shrinkage}
    \frac{\partial [\pmb{\eta}(\widehat{\mathbf{r}})]_{n,t}}{\partial \widehat{r}_{n,t}} =
    \theta_3 \frac{[\mathbf{q}(\widehat{\mathbf{r}}_{n})]_t+\frac{\theta_{1,n}}{\sigma^2(\theta_{1,n}+\sigma^2)}
    ([\mathbf{q}(\widehat{\mathbf{r}}_{n})]_t-1)|\widehat{r}_{n,t}|^2}
    {\Big(1+\frac{\sigma^2}{\theta_{1,n}}\Big)[\mathbf{q}(\widehat{\mathbf{r}}_{n})]_t^2} - \theta_4,
\end{equation}
where the vector function $\mathbf{q}(\widehat{\mathbf{r}}_{n}) = [[\mathbf{q}(\widehat{\mathbf{r}}_{n})]_1,[\mathbf{q}(\widehat{\mathbf{r}}_{n})]_2,\dots,[\mathbf{q}(\widehat{\mathbf{r}}_{n})]_{T_g+1}]^T \in \mathbb{C}^{(T_g+1) \times 1}$ is defined to simplify the expression of (\ref{equ:derivative_sca_SMV_shrinkage}). The function $[\mathbf{q}(\widehat{\mathbf{r}}_{n})]_{t}$ is defined as
\begin{equation}\label{equ:qR}
    [\mathbf{q}(\widehat{\mathbf{r}}_{n})]_t = \frac{\exp\Big(\frac{||\widehat{\mathbf{r}}_{n}||_2^2}{\sigma^2(1+\frac{\sigma^2}{\theta_{1,n}})}\Big) + \big(1+\frac{\theta_{1,n}}{\sigma^2}\big)\theta_{2}}{\exp\Big(\frac{|\widehat{r}_{n,t}|^2}{\sigma^2(1+\frac{\sigma^2}{\theta_{1,n}})}\Big)}.
\end{equation}

\subsection{Parameter training}

\begin{algorithm}[t]
\renewcommand{\algorithmicrequire}{\textbf{Initialize:}}
\renewcommand\algorithmicensure {\textbf{Output:} }
\caption{Parameter training of the tied LAMP network via layer-by-layer and denoiser-by-denoiser learning strategy}\label{Alg:LAMP-SMV_training}
\begin{algorithmic}[1]
\STATE Initialize: $\mathbf{W}=\tilde{\mathbf{S}}^H$ and $\pmb\vartheta_i = \pmb\vartheta^0$, $ i = 1, \dots, I$.
\FOR{$i=1$ to $I$}
\STATE Learn $\{\pmb\vartheta_i\}$ with fixed $\{\mathbf{W}, \{\pmb\vartheta_l\}_{l=1}^{i-1}\}$ with the loss function $\mathcal{L}(\mathbf{\hat{x}}_i) = ||\mathbf{\hat{x}}_i - \mathbf{x}^0||_2^2$.
\STATE Re-learn $\{\{\pmb\vartheta_l\}_{l=1}^{i}\}$ with the loss function $\mathcal{L}(\mathbf{\hat{x}}_i) = ||\mathbf{\hat{x}}_i - \mathbf{x}^0||_2^2$.
\ENDFOR
\STATE Re-learn $\{\mathbf{W}, \{\pmb\vartheta_l\}_{l=1}^{I}\}$ for refinement with the loss function $\mathcal{L}(\mathbf{\hat{x}}_I) = ||\mathbf{\hat{x}}_I - \mathbf{x}^0||_2^2$.
\STATE Return $\{\mathbf{W}, \{\pmb\vartheta_i \}_{i=1}^I\}$.
\end{algorithmic}
\end{algorithm}

In the learnable parameter set of the LAMP network, the matched filter matrices $\mathbf{W}_i$ can be fixed for all layers, i.e., $\mathbf{W}_i = \mathbf{W}$, or vary at each layer $i$. Accordingly, the LAMP network can be referred to as ``tied'' and ``untied'', respectively. Intuitively, the tied LAMP network is a special case of the untied LAMP network, and the untied LAMP network is superior to the tied LAMP network if there is sufficient training data. However, in our simulation trials, we find that the untied neural network brings little performance improvement, and may even make the network prone to overfitting if the training data size is not large enough. Thus, we employ the tied LAMP network in this work.

Apart from the network structure and the parameter set, the training strategy also determines the performance of the neural network. The standard training strategy is the end-to-end training where all the parameters are updated simultaneously by following the back-propagation rule. However, it is found that the LAMP network with end-to-end training can easily converge to a bad local optimal solution due to overfitting. The work \cite{Metzler_2017_NIPS} proposes the layer-by-layer training and the denoiser-by-denoiser training methods to avoid overfitting. In the layer-by-layer training, there are totally $I$ epoches. We first train the learnable parameters of the first layer in the first epoch, then we train the 2-layer sub-network including the first two layers in the second epoch. In the $i$th epoch, the parameters of the $i$-layer sub-network consisting of the first $i$ layers are all trained. The training process repeats until the $I$th epoch is finished.
In specific, at the start of the training process, we initialize the network parameters as stated in line 1 of \textbf{Algorithm} \ref{Alg:LAMP-SMV_training}. Then in each of the following epoches, we firstly decouple the shrinkage functions which recovers $\mathbf{x}^0$ from $\widehat{\mathbf{r}}$ from the target subnetwork to learn its shrinkage parameters $\pmb{\vartheta}_i$ as stated in line 3, which is referred as the denoiser-by-denoiser training. And then all the shrinkage parameters $\{\pmb{\vartheta}_{l}\}_{l=1}^i$ of the target subnetwork in the current epoch are updated simultaneously as stated in line 4. Finally, all the network parameters $\{\mathbf{W}, \{\pmb\vartheta_i \}_{i=1}^I\}$ are all updated for refinement.
The whole training process based on the training strategy that integrates the denoiser-by-denoiser training into the layer-by-layer training approach for the tied LAMP network is outlined in \textbf{Algorithm} \ref{Alg:LAMP-SMV_training}.
It should be mentioned that the number of the training iterations is not predetermined in each epoch. The $i$th epoch finishes until the performance of the updating subnetwork keeps being worse than its best performance achieved in this epoch for a certain number of iterations $T_w$.
It is proven in \cite{Metzler_2017_NIPS} that this training strategy can enable such Learned denoising-based AMP (LDAMP) network to achieve MMSE optimality in theory by state evolution when the following conditions hold:
\begin{itemize}
  \item The measurement matrix $\tilde{\mathbf{S}}$ has i.i.d. sub-Gaussian elements.
  \item The noise $\mathbf{z}$ is i.i.d Gaussian.
  \item The shrinkage functions $\pmb\eta(\cdot)$ are Lipschitz-continuous.
\end{itemize}
Though the elements of the measurement matrix $\tilde{\mathbf{S}}$ are not i.i.d sub-Gaussian, we observe that LAMP can still approach the optimal performance closely from the numerical results.

\vspace{-0.5cm}
\subsection{Discussion}

The LAMP network is classified as the model-driven deep learning category that mixes hand-designed and data-driven methods \cite{He_2019_WCM}. Compared with traditional data-driven deep learning methods which use conventional multi-layer perceptions or convolution networks, there are two main advantages of the neural network LAMP. First, it does not require a very large volume of training data due to its well-designed structure, meaning that the cost in collecting the training data can be reduced. Second, there is no need to re-train the neural network when only the noise variance $\sigma_z$ changes since the noise variance is not used in the AMP algorithm. However, it is mentioned that the LAMP network still needs to be re-trained when the other channel statistics change.

\vspace{-0.4cm}
\section{Deep-Learned AMP in the Multiple-Antenna Scenario}\label{Sec:LAMP-MMV}

\begin{figure}[t]
  \centering
  \subfigure[Distributed LAMP network]
  {\includegraphics[width=.5\textwidth]{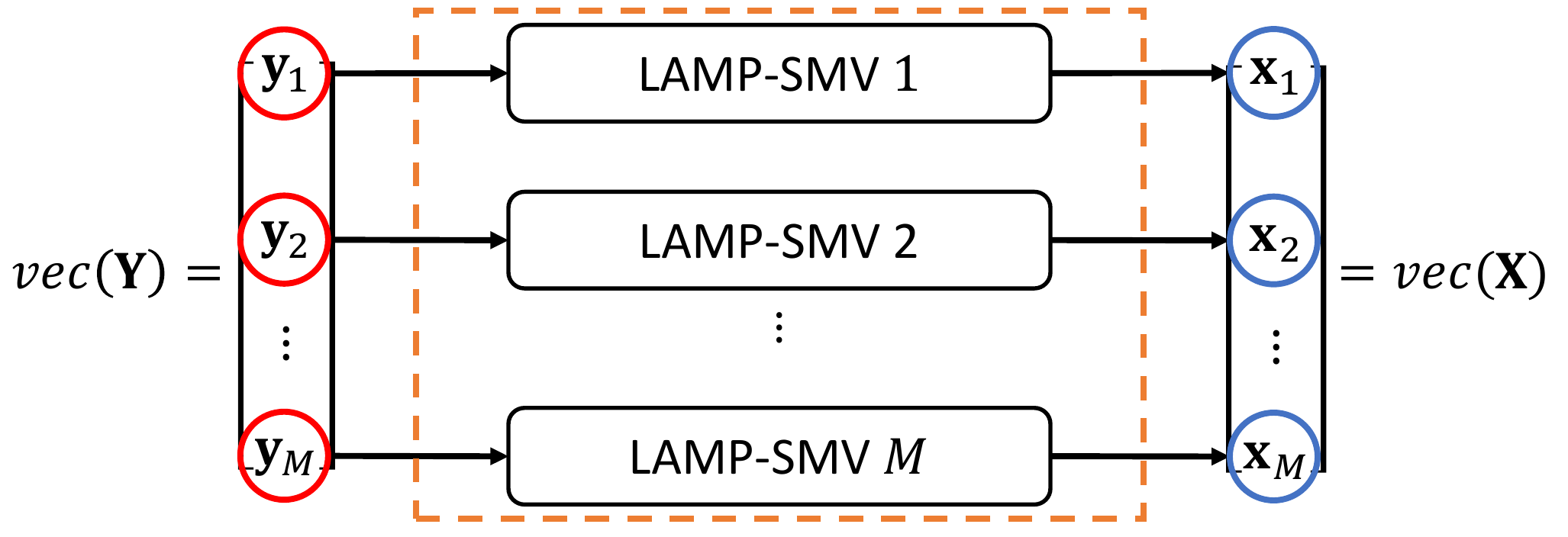}\label{Fig:LAMP-MMV-D}}
  \subfigure[Centralized LAMP network]
  {\includegraphics[width=.5\textwidth]{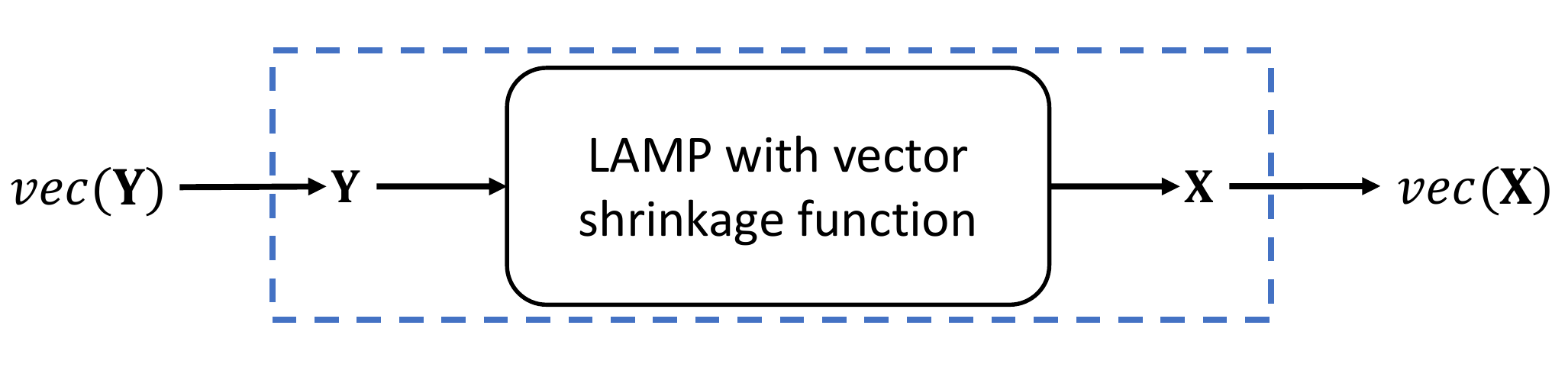}\label{Fig:LAMP-MMV-C}}
  \subfigure[Hybrid LAMP network]
  {\includegraphics[width=.5\textwidth]{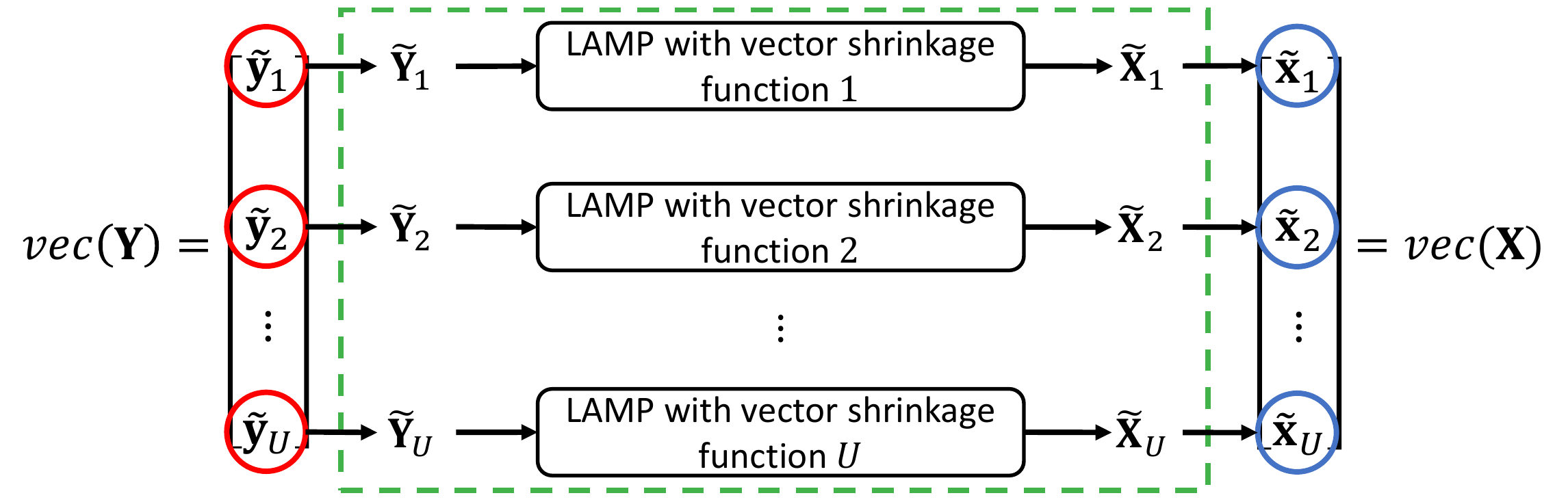}\label{Fig:LAMP-MMV-H}}
  \vspace{-0.5cm}
  \caption{The proposed three LAMP network structures in the multiple-antenna scenario}\label{Fig:LAMP-MMV}
  \vspace{-1cm}
\end{figure}

In this section, we design the LAMP network for the multiple-antenna scenario. Inspired by the AMP algorithms for the MMV problem reviewed in Section \ref{subSec:MMV_problem}, we propose three LAMP network structures, namely, distributed structure, centralized structure and hybrid structure as shown in Fig. \ref{Fig:LAMP-MMV}. The distributed LAMP network (LAMP-D) allows taking advantage of distributed computation units to reduce the running time cost, and the centralized LAMP network (LAMP-C) achieves better recoverability and allows the theoretical performance analysis by using the state evolution. The hybrid LAMP network (LAMP-H) combines the advantages of both the distributed network and the centralized network. The details are presented in the following subsections. In the rest of paper, to avoid confusion, we denote the LAMP network designed in the previous section for the single-antenna case as LAMP-SMV.


\vspace{-0.4cm}
\subsection{Distributed LAMP Network}

The structure of the LAMP-D network is shown in Fig. \ref{Fig:LAMP-MMV-D}, which is comprised by $M$ parallel LAMP-SMV subnetworks.
The parallel AMP-MMV algorithm adds the extra operations of belief refinement and exchanges to improve the recoverability after solving the $M$ separate SMV problems by AMP in (\ref{Alg:AMP-SMV_iteration}). However, the channel statistics are still needed in the AMP estimation for each SMV problem. By employing $M$ independent AMP-SMV solvers with known channel distributions and user active probability, we find the MMV problem can also be well solved.
Thus, we propose to construct the LAMP-D network by using $M$ independent LAMP-SMV subnetworks to estimate the channel coefficients on their corresponding antennas in parallel.

In the construction of the LAMP-D network, whether the $M$ LAMP-SMV networks share the same parameter value or have different parameter values needs to be studied firstly. When the fading coefficient on each antenna at the BS is i.i.d., all the LAMP-SMV subnetworks can share the same parameters value. Therefore, only one LAMP-SMV network is trained in practice and then the learned parameter values are shared by all LAMP-SMV subnetworks. The received signal and the effective channel are denoted as $\mathbf{Y} = [\mathbf{y}_1, \mathbf{y}_2, \dots, \mathbf{y}_M]$ and $\mathbf{X} = [\mathbf{x}_1, \mathbf{x}_2, \dots, \mathbf{x}_M]$, and each $(\mathbf{y}_m, \mathbf{x}_m)$ is considered to be one sample for training. So that the quantity of the training data for the neural network becomes $MD_T$, where $D_T$ is the number of received signals $\mathbf{Y}$ in the training data set.
When the fading coefficient on each antenna is not i.i.d., there are $M$ independent LAMP-SMV subnetworks to be trained. Each subnetwork needs to be trained based on the received signals on the corresponding antenna, and then the size of training data set for each subnetwork is only $D$. In this paper, the neural network with all LAMP-SMV subnetworks sharing the same parameter value is adopted, since the i.i.d. channel distribution on each antenna is assumed and the LAMP-SMV subnetwork can be trained by more training data samples to achieve better recovery performance. The LAMP-SMV subnetworks in the LAMP-D network are trained by following \textbf{Algorithm} \ref{Alg:LAMP-SMV_training}.
It is concluded that the LAMP-D network concentrates on improving the recoverability on each single antenna and the distributed setup of subnetworks can reduce running time cost by parallel computation. While the optimal performance of the AMP framework may not be achieved, since the common sparsity in the estimated channel matrix is not fully exploited.

\subsection{Centralized LAMP Network}

\begin{figure}[t]
  \centering
  \includegraphics[width=.5\textwidth]{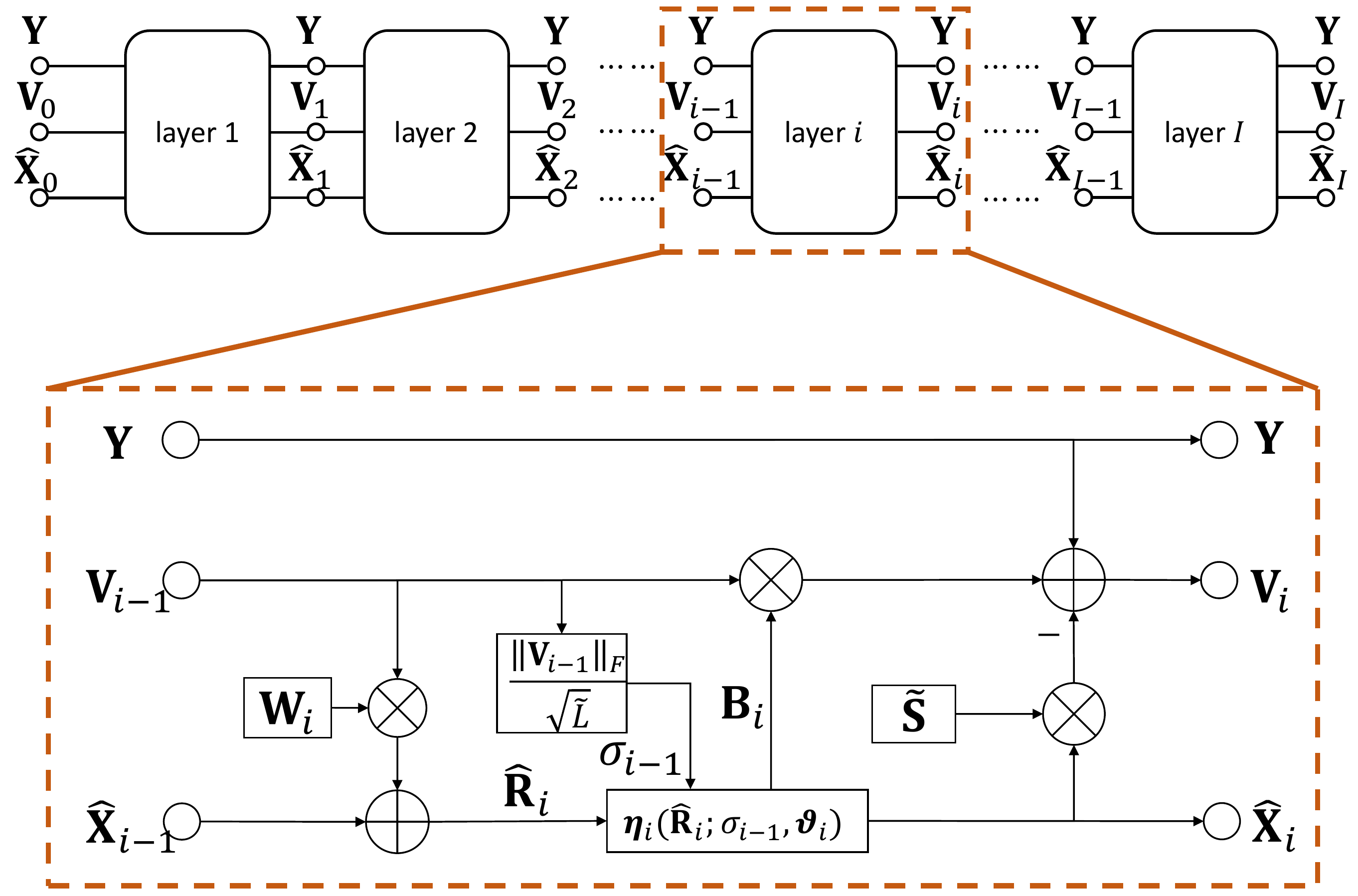}
  \vspace{-0.2cm}
  \caption{The network structure and $i$th layer of the LAMP network with vector shrinkage function}\label{Fig_LAMP-MMV-C_layer}
  \vspace{-1cm}
\end{figure}

The centralized LAMP network contains only one single LAMP network with vector shrinkage function which ``unfolds'' the iterations of the AMP algorithm with vector shrinkage function. The vector shrinkage function usually recovers each row $\mathbf{x}^0_{n,t}$ in the true signal $\mathbf{X}^0$ individually to exploit the common sparsity in the columns of $\mathbf{X}^0$, so that the LAMP-C network can obtain a more accurate estimate of $\mathbf{X}^0$ than the LAMP-D network. The network structure and the details of the $i$th layer in the LAMP network with vector shrinkage function is similar to LAMP-SMV from Fig. \ref{Fig_LAMP-MMV-C_layer}. In the neural network, the input and output are usually limited to be in the form of a vector, while the received signal and the estimated channel are both in the form of a matrix. Thus, a simple form transformation operation is needed at the input and output of the network which is eliminated in Fig. \ref{Fig_LAMP-MMV-C_layer}. In particular, this transformation operation also needs to be added both in the input and the output of each layer, since the layer-by-layer and denoiser-by-denoiser training strategy is also adopted to train the LAMP-C network. In addition, we also select the MSE function as the loss function of the LAMP-C network, i.e., $\mathcal{L}(\widehat{\mathbf{X}}) = ||vec(\widehat{\mathbf{X}}) - vec(\mathbf{X}^0)||_2^2 = ||\widehat{\mathbf{X}} - \mathbf{X}||_F^2$. We can also represent the learnable parameter set as $\{\mathbf{W}_i, \pmb{\vartheta}_i\}_{i=1}^I$, where $\mathbf{W}_i$ is regarded as the matched filter matrix and $\pmb{\vartheta}_i$ is the learnable parameters set of the vector shrinkage functions in the $i$th layer. We drop the index $i$ to simplify the expression in the following part. In the LAMP network with vector shrinkage function, the ST function and the MMSE-optimal denoising function are also adopted. The ST function can be presented as
\begin{equation}\label{fun:st_MMV}
    [\pmb{\eta}(\widehat{\mathbf{R}})]_{n,t} = \left( \widehat{\mathbf{r}}_{n,t} - \theta\sqrt{M}\sigma \frac{\widehat{\mathbf{r}}_{n,t}}{||\widehat{\mathbf{r}}_{n,t}||_2} \right) \cdot \mathbb{I}(||\widehat{\mathbf{r}}_{n,t}||_2 >\theta \sqrt{M}\sigma),
\end{equation}
where $\theta$ is the learnable tuning parameter and the learnable parameter set is  $\pmb\vartheta = \{ \theta \}$. By exploiting the system statistics, the MMSE-optimal denoising function can achieve better recovery performance.
The joint distribution of the effective channel $\mathbf{X}_n$ of user $n$ is given as
\begin{equation}\label{equ:X_BG}
    p(\mathbf{X}_{n}) = (1-p_{a})\prod_{t=1}^{T_g+1}\delta(\mathbf{x}_{n,t}) + p_{a,T}\sum_{t=1}^{T_g+1}\Big(\mathcal{CN}(\mathbf{x}_{n,t};0,\phi_n\mathbf{I}_M)\prod_{t'\ne t}\delta(\mathbf{x}_{n,t'})\Big).
\end{equation}

Based on the MMSE-optimal criteria, the vector shrinkage function can be obtained, which estimates each effective channel matrix $\mathbf{X}_{n}$ rather than treat each $\mathbf{x}_{n,t}$ independently. The covariance matrix of the corrupted noise in $\mathbf{r}_{n,t}$ is a diagonal matrix with identical diagonal elements, then the MMSE-optimal vector shrinkage function can be simplified.
Since the Section \ref{sec:LAMP_SMV} has illustrated how to obtain the learnable shrinkage function based on the MMSE-optimal shrinkage function, we here directly give the learnable MMSE-optimal denoising function as
\begin{equation}\label{equ:vec_shrinkage_learn}
    [\pmb{\eta}(\widehat{\mathbf{R}})]_{n,t} = \theta_3 \frac{\exp\bigg(\frac{||\widehat{\mathbf{r}}_{n,t}||_2^2}{\sigma^2(1+\frac{\sigma^2}{\theta_{1,n}})}\bigg) \widehat{\mathbf{r}}_{n,t} }{(1+\frac{\sigma^2}{\theta_{n}}) \left[\exp\bigg(\frac{||\widehat{\mathbf{R}}_{n}||_F^2}{\sigma^2(1+\frac{\sigma^2}{\theta_{1,n}})}\bigg) + \big(1+\frac{\theta_{1,n}}{\sigma^2}\big)^M \theta_{2}\right]} - \theta_4 \widehat{\mathbf{r}}_{n,t},
\end{equation}
where the learnable parameter set is also defined as $\pmb\vartheta = \{ \{\theta_{1,n}\}_{n=1}^{N}, \theta_{2}, \theta_3, \theta_4\}$ with $\theta_{1,n} = \phi_n$ and $\theta_2 = \frac{1-p_{a}}{p_{a,T}}$.
Also, we can set $\theta_{1,n} = \theta_1, \forall n=1,\dots,N$ when all users suffer the same large-scale channel attenuation.
In each layer of the LAMP-C network, the Jacobi matrix of the vector shrinkage function also needs to be calculated to obtain the Onsager term, which can be written as
\begin{align}\label{equ:derivative_vec_shrinkage_learn}
    \frac{\partial [\pmb{\eta}(\widehat{\mathbf{R}})]_{n,t}}{\partial \widehat{\mathbf{r}}_{n,t}} =
    \theta_3 \frac{[\mathbf{Q}(\widehat{\mathbf{R}}_{n})]_t \mathbf{I}_M + \frac{\theta_{1,n}}{\sigma^2(\theta_{1,n}+\sigma^2)}
    ([\mathbf{Q}(\widehat{\mathbf{R}}_{n})]_t-1)\widehat{\mathbf{r}}_{n,t}^H\widehat{\mathbf{r}}_{n,t}}
    {\Big(1+\frac{\sigma^2}{\theta_{1,n}}\Big)[\mathbf{Q}(\widehat{\mathbf{R}}_{n})]_t^2} - \theta_4 \mathbf{I}_M,
\end{align}
where $\mathbf{Q}(\mathbf{R}_n) = [[\mathbf{Q}(\mathbf{R}_n)]_1,[\mathbf{Q}(\mathbf{R}_n)]_2,\dots,[\mathbf{Q}(\mathbf{R}_n)]_{T_g+1}]^T \in \mathbb{C}^{(T_g+1) \times 1}$ is defined for simplifying the expression of (\ref{equ:derivative_vec_shrinkage_learn}) and $[\mathbf{Q}(\mathbf{R}_n)]_t$ is given as
\begin{equation}\label{equ:qR}
    [\mathbf{Q}(\widehat{\mathbf{R}}_{n})]_t = \frac{\exp\Big(\frac{||\widehat{\mathbf{R}}_{n}||_F^2}{\sigma^2(1+\frac{\sigma^2}{\theta_{1,n}})}\Big) + \big(1+\frac{\theta_{1,n}}{\sigma^2}\big) \theta_{2}}{\exp\Big(\frac{|\widehat{\mathbf{r}}_{n,t}|^2}{\sigma^2(1+\frac{\sigma^2}{\theta_{1,n}})}\Big)}.
\end{equation}

The signal flow graph of the LAMP-C network is the same as the AMP algorithm with matrix shrinkage function. Thus, the denoising performance of each layer in the LAMP-C network can also be described by state evolution $\pmb\Sigma_{i}$ in the asymptotic region. Since the layer-by-layer and denoiser-by-denoiser training strategy is proven to enable the LDAMP network \cite{Metzler_2017_NIPS} for the SMV problem to achieve MMSE optimality under the conditions mentioned in Section \ref{sec:LAMP_SMV}, it is reasonable to speculate that the strategy can also ensure the MMSE-optimality of the LAMP-C network under the same conditions. To prove the property, we first follow \cite{Metzler_2017_NIPS} to define a set of variables $\{\tau_i\}_{i=1}^I$ as
\begin{equation}\label{equ:tau}
    \tau_i = \frac{1}{\tilde{N}M}\mathbb{E}_{\mathbf{D}}[||\pmb\eta^i_{\pmb\vartheta_i,\delta_i}(\mathbf{X}^0+\delta_{i-1}\mathbf{D}) - \mathbf{X}^0||_2^2].
\end{equation}
In addition, we also give the definition of the monotone denoising function that $\inf_{\pmb\vartheta}\mathbb{E}_{\mathbf{D}}||\pmb\eta_{\pmb\vartheta,\delta}(\mathbf{X}^0+\delta\mathbf{D})-\mathbf{X}^0||_F^2$ is a non-decreasing function of $\delta$ for any $\mathbf{X}^0$. With the above definition, we are ready to present the following lemma.

\begin{lemma}\label{lemma:LDT_MMSE_opt}
Suppose that the shrinkage functions $\pmb\eta^i_{\pmb\vartheta_i}(\cdot),~i = 1,\dots,I$, are the monotone denoising functions. By following the greedy selection strategy in \cite{Metzler_2016_TIT}, the parameters $\pmb\vartheta_1$ are updated as $\pmb{\vartheta}_1^*$ to minimize $\mathbb{E}_{\mathbf{X}^0}[\tau_1]$ and fixed; then the parameters $\pmb\vartheta_2$ are updated as $\pmb{\vartheta}_2^*$ to minimize $\mathbb{E}_{\mathbf{X}^0}[\tau_2]$ and fixed, ..., and the parameters $\pmb\vartheta_I$ are updated as $\pmb{\vartheta}_I^*$ to minimize $\mathbb{E}_{\mathbf{X}^0}[\tau_I]$. The LAMP network with all the updated learnable parameters $\{\vartheta_i^*\}_{i=1}^{I}$ finally minimizes $\mathbb{E}_{\mathbf{X}^0}[\tau_I]$.
\end{lemma}

\begin{IEEEproof}
This lemma can be proved by replacing $\tau_i$ with $\mathbb{E}_{\mathbf{X}^0}[\tau_i]$ in the proof for Lemma 3 in \cite{Metzler_2016_TIT}.
\end{IEEEproof}

If the conditions in Lemma \ref{lemma:LDT_MMSE_opt} and Section \ref{sec:LAMP_SMV} are all satisfied, the LAMP-C network under layer-by-layer and denoiser-by-denoiser training strategy achieves MMSE optimality. However, the measurement matrix $\tilde{\mathbf{S}}$ has non i.i.d elements, which makes no optimality guarantee in theory. From the numerical results, we can see that the LAMP-C network based on the layer-by-layer and denoiser-by-denoiser training strategy can still approach the optimal performance.

\begin{algorithm}[t]
\renewcommand{\algorithmicrequire}{\textbf{Initialize:}}
\renewcommand\algorithmicensure {\textbf{Output:} }
\caption{Parameter training of the tied LAMP network with vector shrinkage function via layer-by-layer and denoiser-by-denoiser learning strategy}\label{Alg:LAMP-MMV_training}
\begin{algorithmic}[1]
\STATE Initialize: $\mathbf{W}=\tilde{\mathbf{S}}^H$ and $\pmb\vartheta_i = \pmb\vartheta^0$, $  = 1, \dots, I$.
\FOR{$i=1$ to $I$}
\STATE Learn $\{\pmb\vartheta_i\}$ with fixed $\{\mathbf{W}, \{\pmb\vartheta_l\}_{l=1}^{i-1}\}$ based on the loss function $\mathcal{L}(\mathbf{\widehat{X}}_i) = ||\mathbf{\widehat{X}}_i - \mathbf{X}^0||_F^2$.
\STATE Re-learn $\{\{\pmb\vartheta_l\}_{l=1}^{i}\}$ based on the loss function $\mathcal{L}(\mathbf{\widehat{X}}_i) = ||\mathbf{\widehat{X}}_i - \mathbf{X}^0||_F^2$.
\ENDFOR
\STATE Re-learn $\{\mathbf{W}, \{\pmb\vartheta_i\}_{i=1}^{I}\}$ for refinement based on the loss function $\mathcal{L}(\mathbf{\widehat{X}}_I) = ||\mathbf{\widehat{X}}_I - \mathbf{X}^0||_F^2$.
\STATE Return $\{\mathbf{W}, \{\pmb\vartheta_i \}_{i=1}^I\}$.
\end{algorithmic}
\end{algorithm}

Compared with the LAMP-SMV network, the learnable parameters set $\{\mathbf{W}_i, \pmb{\vartheta}_i\}_{i=1}^I$ of the proposed LAMP-C network has no difference. And the LAMP-C network can be similarly divided by the tied version and the untied version depending on whether the matched filter matrix $\mathbf{W}_i$ for $i=1, \dots, I$, share the same values in all layers or have different values across different layers. Therefore, the training process of the LAMP-C network will also keep nearly unchanged. We outline the training procedure of the tied LAMP-C network in \textbf{Algorithm} \ref{Alg:LAMP-MMV_training}.

\subsection{Hybrid LAMP Network}

The neural networks proposed in the above two subsections solve the MMV problem in either purely distributed way or centralized way, which concentrates only on reducing the running time cost or improving the recovery accuracy in the multiple-antenna scenario. Intuitively, we can construct the LAMP network in a hybrid way, which has combined the structures of both the LAMP-D network and the LAMP-C network to save the running time and improve the recoverability simultaneously. The structure of the LAMP-H network is shown in Fig. \ref{Fig:LAMP-MMV-H} that consists of $U$ parallel LAMP networks with vector shrinkage function, and each LAMP subnetwork only recovers the channel on its corresponding subset of antennas. When the channel coefficient on each antenna is assumed to satisfy i.i.d. distribution, we can divide all antennas into $U$ non-overlap subsets with the same size by sequence. Thus, the effective channel matrix can be represented by $\mathbf{Y} = [\tilde{\mathbf{Y}}_1, \dots, \tilde{\mathbf{Y}}_U]$ where $\tilde{\mathbf{Y}}_u \in \mathbb{R}^{L \times \frac{M}{U}}$ and we assume that $\frac{M}{U}$ is an integer here. Then the input vector of the $u$st LAMP subnetwork can be denoted as $\tilde{\mathbf{y}}_{u}$, where $\tilde{\mathbf{y}}_u = vec(\tilde{\mathbf{Y}}_u)$. And the output estimated channel matrix $\widehat{\mathbf{X}}$ can also be obtained by concatenating the estimated sub-channels $\{\widehat{\mathbf{X}}_u\}_{u=1}^U$, i.e., $\widehat{\mathbf{X}} = [\widehat{\mathbf{X}}_1, \dots, \widehat{\mathbf{X}}_U]$.
Similar to the LAMP-D network, the learnable parameters of different subnetworks in LAMP-H can also have different value setting according to the considered system. On the other hand, it is also implied that the antennas can be divided into different subsets in various ways, and the number of antennas in different subsets can even be various in some cases. However, the best dividing method is unknown which usually needs many trials before the neural network deployment. In this work, the LAMP-H network is designed to have the same parameter value in each LAMP subnetwork with vector shrinkage function. Each LAMP subnetwork estimates the channel coefficients on the corresponding $\frac{M}{U}$ antennas, and the size of the training data for the subnetwork training is $\frac{MD}{U}$.

\vspace{-0.2cm}
\subsection{Discussion}

\begin{table*}[t]
  \centering
  \caption{Computational Complexity of Different Algorithms in the Single-Antenna Scenario}
  \label{table:CCCSA}
  \vspace{-0.3cm}
    \begin{tabular}{c|cccc}
    \hline
    \hline
    Algorithm   & BPDN & OMP & AMP & LAMP \\
    \hline
    Complexity & $\mathcal{O}(\tilde{N}^{3.5})$ & $\mathcal{O}(IK^3)$ & $\mathcal{O}(I\tilde{L}\tilde{N})$ & $\mathcal{O}(I\tilde{L}\tilde{N})$ \\
    \hline
	\end{tabular}
\vspace{-0.5cm}
\end{table*}

\begin{table*}[t]
  \centering
  \caption{Computational Complexity of the Proposed LAMP networks in the Multiple-Antenna Scenario}
  \label{table:CCCMA}
  \vspace{-0.3cm}
    \begin{tabular}{c|ccc}
    \hline
    \hline
    Network & LAMP-D & LAMP-C & LAMP-H \\
    \hline
    Complexity & $\mathcal{O}(I\tilde{L}(\tilde{N}M+M))$ & $\mathcal{O}(I\tilde{L}(\tilde{N}M+\tilde{L}M^2))$ & $\mathcal{O}(I\tilde{L}(\tilde{N}M+\frac{M^2}{U}))$ \\
    \hline
	\end{tabular}
\vspace{-0.5cm}
\end{table*}

Since the AMP algorithm with vector shrinkage function has saturated recovery performance when $M$ exceeds a certain value, the performance the LAMP-C network will be also limited in such case, which has been validated by the numerical results.
Therefore, the LAMP-H network usually can be a better choice which can approach the optimal performance and make use of the distributed computing units to save the running time simultaneously when the number of antennas is very large.

\subsection{Computational Complexity Analysis}

To implement the proposed algorithms, the computational complexity also has a great influence to the hardware usage and power consumption. Here, we analyze the computational complexity of our proposed algorithm and compare them with the conventional CS algorithms. Table \ref{table:CCCSA} gives the complexities of the proposed LAMP network as well as that of the conventional CS algorithm BPDN, OMP, and the AMP algorithm in the single-antenna scenario. And Table \ref{table:CCCMA} lists the complexities of the three LAMP networks in the multiple-antenna scenario. The number of iterations or the layers of network is denoted as $I$ in these two tables. By comparison, the complexity of the AMP algorithm and the LAMP networks increases linearly with $L$, $N$, since no matrix inversion is needed. In Table \ref{table:CCCMA}, the complexity of LAMP-C and LAMP-H networks increase linearly with $M^2$, while LAMP-D network increase linearly with $M$. Additionally, when perfect system statistics are unavailable, more well-performed shrinkage functions can be utilized in LAMP, which can also speed up the convergence of the AMP framework to reduce the needed iterations. Thus, the LAMP network is more computationally efficient in massive access.

\section{Simulation Results}\label{Sec:simulation}

\begin{figure}[t]
  \centering
  \includegraphics[width=.47\textwidth]{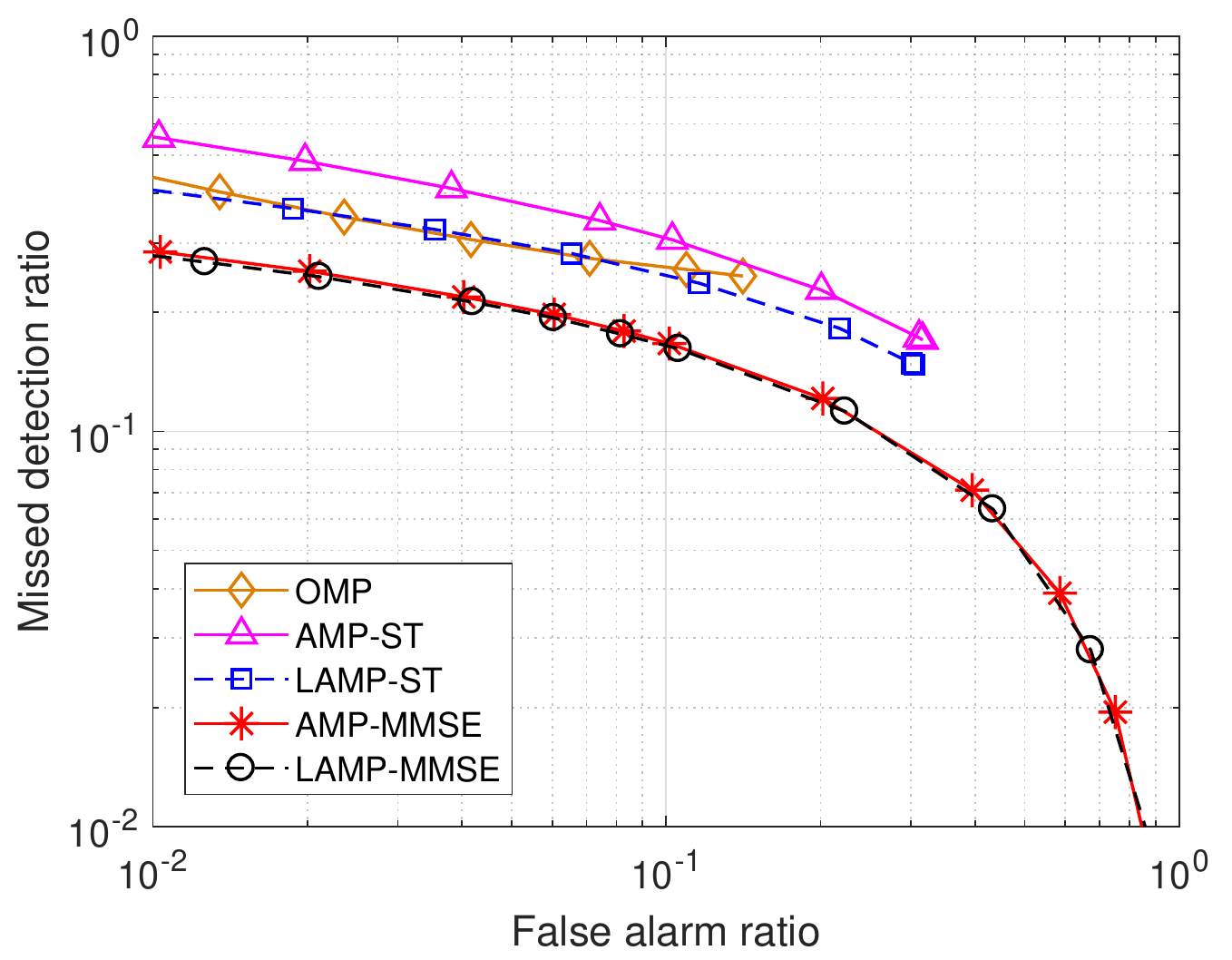}
  \vspace{-0.4cm}
  \caption{The user activity detection performance comparison in the single-antenna scenario with $T_g=3$.}\label{Fig:SMV_Pm-Pf}
  \vspace{-0.8cm}
\end{figure}

\begin{figure}[t]
  \centering
  \includegraphics[width=.47\textwidth]{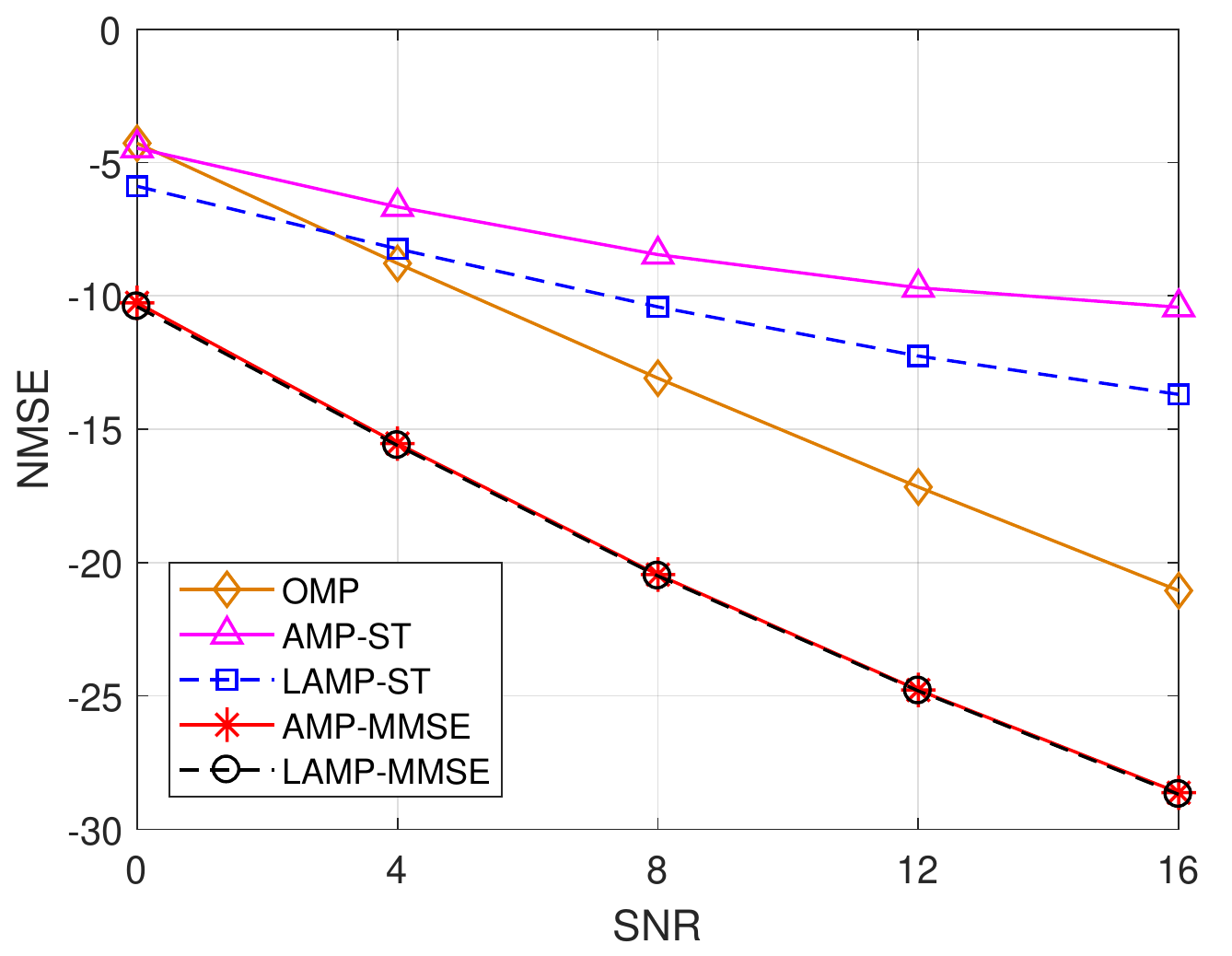}
  \vspace{-0.4cm}
  \caption{The channel estimation performance comparison in the single-antenna scenario with $T_g=3$.}\label{Fig:SMV_NMSE}
  \vspace{-0.8cm}
\end{figure}

In this section, we present the simulation results of the proposed algorithms in asynchronous grant-free random access systems.

We consider a system with $N=100$ users for illustration purpose, although the AMP-based algorithms can be used for a much larger-scale problem. Each user has a probability of $p_a=0.1$ to be active.
To simplify the demonstration, all users are placed at the edge of the cell centered by the BS. Thus, the large-scale channel attenuation $\phi_n$ of all users are equal, i.e., $\phi_n = \phi, \forall n$ and the  signal-to-noise ratio is defined by $\text{SNR} = \frac{\phi}{\sigma_z^2}$. We consider $\text{SNR} = 0\text{dB}$ in the simulations if not specified otherwise.
We set the pilot length $L=40$, 
and the maximal symbol delay of users and the length of guide time are set to be $3$, i.e., $T_g=D=3$.
The small-scale fading coefficient of each user at each antenna is generated according to the i.i.d. Gaussian distribution with zero mean and unit variance, i.e., $g_{n,m} \sim \mathcal{CN}(0,1)$, $\forall n, m$.
We generate $10^5$ independent samples with $\text{SNR}=0\text{dB}$ for training and $5\times10^3$ independent samples with the same distribution for validation in both the single-antenna scenario and the multiple-antenna scenario. Then another $5\times10^3$ independent samples are generated for each $\text{SNR} (\text{dB}) \in \{0,4,8,12,16\}$, which leads to totally $2.5 \times 10^4$ testing samples.
In the training process, the training data is divided into minibatches of size $100$.
The performance of the neural network on the test set is also evaluated by MSE of all the samples in this set, which is computed as $\text{MSE} = \frac{1}{B}\sum_{b=1}^{B}||\widehat{\mathbf{x}}_{b}-\mathbf{x}_{b}||_2^2$ with $B$ being the number of samples.
The number of layers in the LAMP networks for both the single-antenna scenario and multiple-antenna scenario is set to be $I = 10$.\footnote{Based on our simulation trials, the AMP algorithm will converge in 10 iterations in our system setting, so that we also set the number of the layers in the LAMP network as $I=10$.}
The neural networks are all trained and tested by using the deep neural network library TensorFlow. The Adam optimizer is adopted with a training rate $7\times10^{-4}$.
The AMP algorithm and the OMP algorithm are employed as the benchmarks which are also evaluated on the same test data set.



\begin{figure}[t]
  \centering
  \includegraphics[width=.47\textwidth]{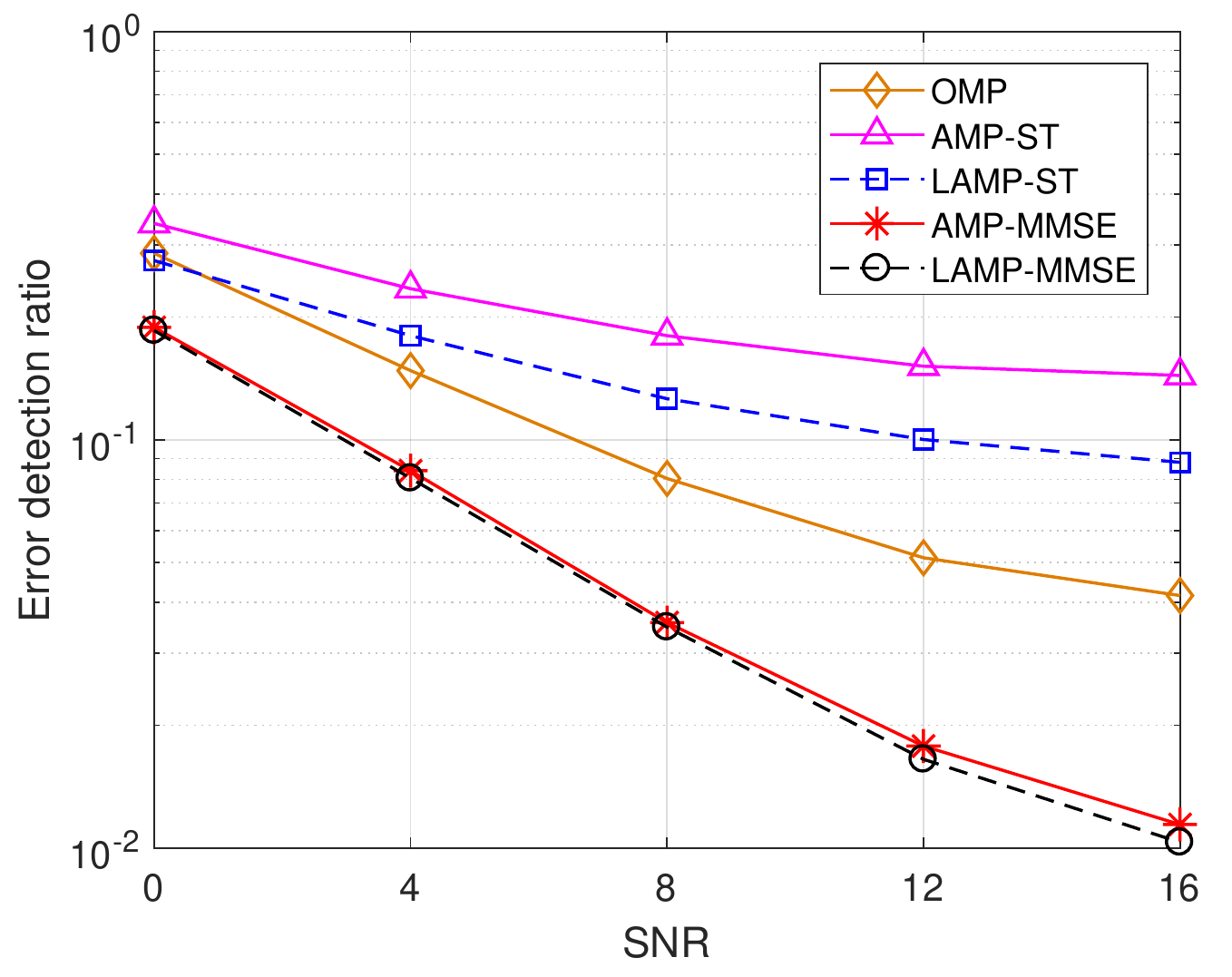}
  \vspace{-0.4cm}
  \caption{The delay detection performance comparison in the single-antenna scenario with $T_g = 3$ under fixed false alarm ratio $\epsilon = 0.1$.}\label{Fig:ER_Tg}
  \vspace{-1cm}
\end{figure}

In Fig. \ref{Fig:SMV_Pm-Pf}, we first evaluate the active user detection performance of the LAMP network in terms of the missed detection ratio versus the false alarm ratio by varying the decision threshold value $q_{th}$.
The missed detection ratio is defined as the ratio of the number of undetected users over the number of active users, and the false alarm ratio is defined as the ratio of the number of the inactive users falsely detected to be active over the number of the inactive users.
Here, user $n$ is detected to be active with symbol delay $t_{max}$ if the element with maximal magnitude, denoted as $\widehat{x}_{n,t_{max}+1}$, in $\widehat{\mathbf{x}}_n$ is larger than $q_{th}$. AMP-ST and LAMP-ST denote the schemes with the ST function being used, while AMP-MMSE and LAMP-MMSE employ the MMSE-optimal denoising function. When the MMSE-optimal denoising function is selected, the exact value of the parameters is unknown and will be learned in LAMP while it is perfectly known in AMP.
Since the tuning parameter $\theta$ of ST function is fixed in the AMP-ST and LAMP-ST, many elements in the output vector of the ST function are zeros. This means that the corresponding users will never be detected to be active, thus the false alarm ratio of the curves of AMP-ST and LAMP-ST cannot approach one. In this paper, only the $2p_aN$ users which are most likely to be active are detected by the OMP algorithm, so that its false alarm ratio also never approaches one.
We observe that LAMP-ST significantly outperforms AMP-ST by learning the optimal tuning parameter value and the matched filter matrix. It achieves similar performance to the OMP algorithm, since LAMP-ST implicitly solves the LASSO problem and approaches the performance of LASSO in the asymptotic regime. And the learnable MMSE-optimal shrinkage function can enable LAMP to outperform both LAMP-ST and OMP by exploiting the statistical information of the system. It is observed that the LAMP-MMSE network can slightly outperform the AMP-MMSE algorithm, which benefits from the deep learning techniques to learn the optimal network parameters.

Fig. \ref{Fig:SMV_NMSE} shows the channel estimation performance in terms of the normalized mean square error (NMSE) versus SNR, where the metric is defined as $\text{NMSE} = \frac{||\hat{\mathbf{h}}-\mathbf{h}^0||_2^2}{||\mathbf{h}^0||_2^2}$.
Similar to the user activity performance, the DL techniques enable the LAMP network with ST function to achieve lower channel estimation error than AMP-ST. The LAMP-ST network is observed to outperform the OMP algorithm at $\text{SNR}=0\text{dB}$, but the OMP algorithm can achieve much lower NMSE than the LAMP-ST network with SNR increasing, which implies that the least square estimation can offer better recoverability in high SNR regime.
Compared with the ST function, the MMSE-optimal denoising function can provide significant performance improvement for the AMP framework when SNR is larger, since the system statistics are exploited.

\begin{figure}[t]
  \centering
  \includegraphics[width=.47\textwidth]{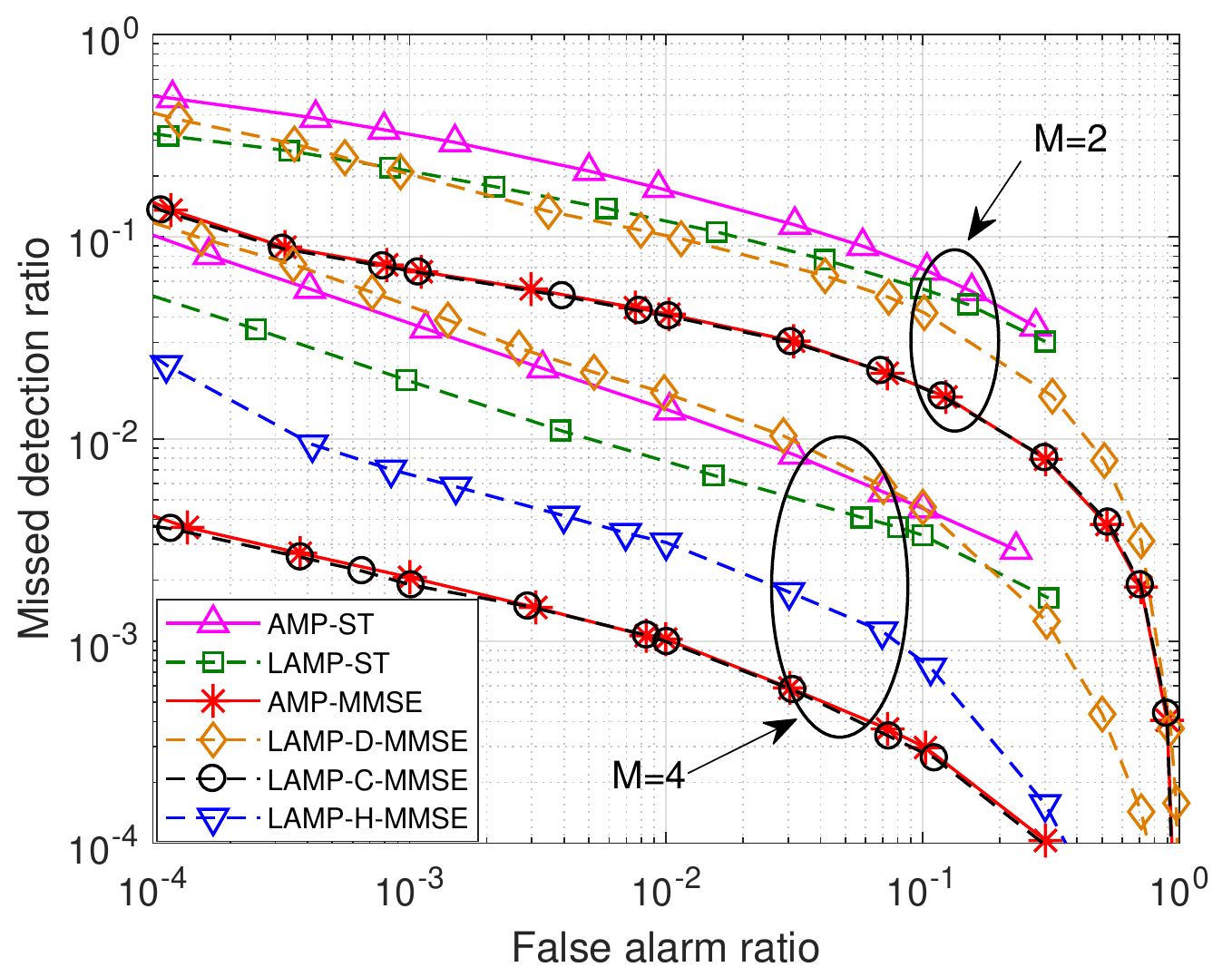}
  \vspace{-0.4cm}
  \caption{The user activity detection performance in the multiple-antenna scenario with $T_g=3$.}\label{Fig:MMV_Pm-Pf}
  \vspace{-1cm}
\end{figure}

The delay detection performance is shown in Fig. \ref{Fig:ER_Tg}. Here, the error detection ratio is defined by the ratio of the number of the active users with wrongly detected delay to the number of active users. The error detection ratios are obtained under a fixed false alarm ratio $\epsilon=0.1$.
Since the delays of the active users are detected based on the estimated channel $\widehat{\mathbf{x}}$ or $\widehat{\mathbf{X}}$, the delay detection performance is heavily influenced by the channel estimation performance.
Thus, we can find that the delays are usually detected more accurately when more precise channel estimation is achieved.
In particular, the LAMP-MMSE network can achieve slightly larger error detection ratio reduction than the AMP-MMSE algorithm with SNR increasing, which also inversely implies that LAMP-MMSE can enable slightly better channel estimation of the active users in the large SNR scenario.

In the following, we consider the scenario where the BS has multiple antennas. The ST function is only employed in the LAMP-C networks, and the MMSE-optimal denoising function is adopted in all three kinds of LAMP networks. The LAMP-H network evaluated here consists of two LAMP networks with vector shrinkage function that estimate the channels at two antennas when $M=4$.
We first evaluate the user activity detection performance shown in Fig. \ref{Fig:MMV_Pm-Pf}.
It is observed that increasing $M$ can dramatically improve the user activity detection performance of the LAMP networks. It is shown that the centralized network always outperforms decentralized network and the hybrid network when the type of the shrinkage function and $M$ are constant, which indicates that exploiting the common sparsity at all antennas can provide significant performance improvement in user activity detection.
In particular, though the statistical information is not fully exploited in the LAMP-ST network, it can outperform LAMP-D-MMSE when $M = 4$. This result implies that centralized structure is potential to provide more performance improvement than the well-designed shrinkage function by increasing the number of antennas at the BS. The performance of LAMP-H-MMSE lies between that of LAMP-D-MMSE and LAMP-C-MMSE, since it balances the computational complexity and the recoverability.


\begin{figure}[t]
  \centering
  \subfigure[$M=2$]
  {\includegraphics[width=.42\textwidth]{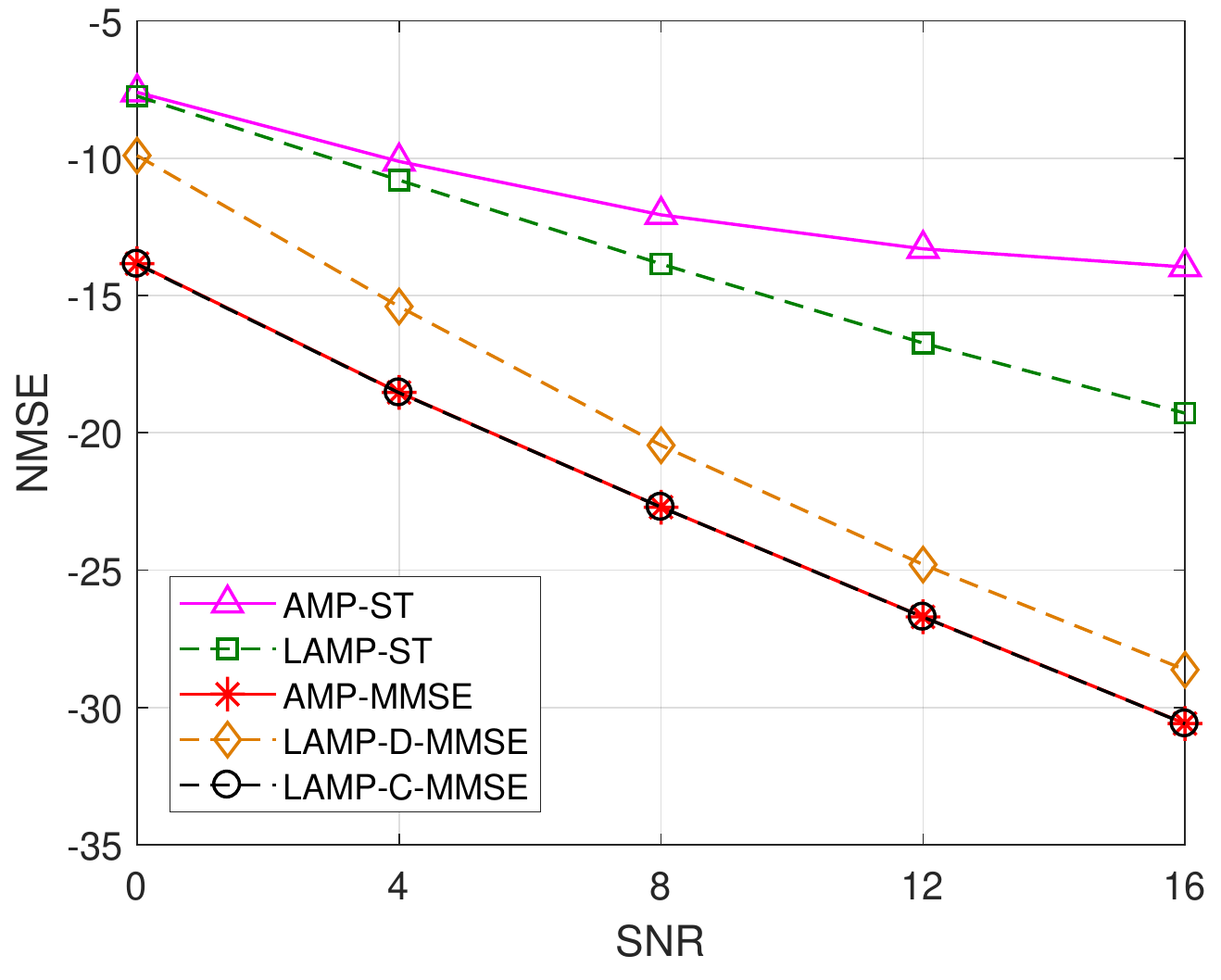}\label{Fig:MMV_NMSE_E2}}
  \subfigure[$M=4$]
  {\includegraphics[width=.42\textwidth]{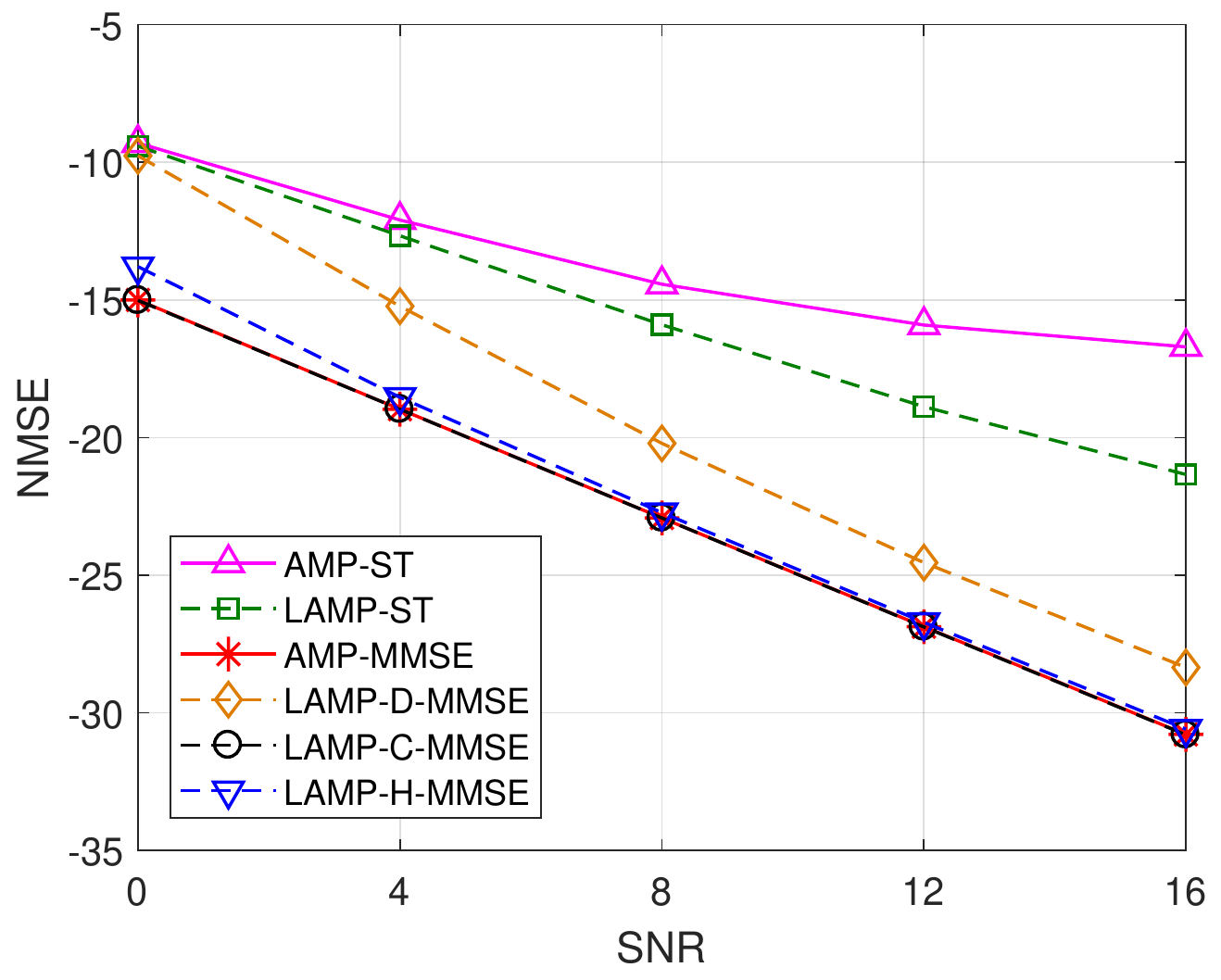}\label{Fig:MMV_NMSE_E4}}
  \caption{The channel estimation performance in the multiple-antenna scenario with $T_g=3$.}\label{Fig:MMV_NMSE}
  \vspace{-0.8cm}
\end{figure}

Fig. \ref{Fig:MMV_NMSE} shows the channel estimation performance of the LAMP networks in the multiple-antenna scenario. The metric is defined as $\text{NMSE} = \frac{\hat{||\mathbf{H}} - \mathbf{H}^0||^2_F}{||\mathbf{H}^0||^2_F}$. 
We can see that the channel estimation performance of the LAMP-D network keeps almost unchanged when the $M$ increases. This implies that the size of our training data is large enough to enable the LAMP network work well, which is usually unable to train a traditional deep neural network well.
It is also show that the performance gap between LAMP-C-MMSE and the other LAMP networks including LAMP-D-MMSE and LAMP-H-MMSE will become smaller with SNR increasing. And the NMSE of LAMP-H-MMSE will approach that of LAMP-C-MMSE in the high SNR regime. These results all imply that the LAMP-C-MMSE may have saturated performance when $M$ or SNR is large enough due to the limit of the AMP framework.
Thus, the LAMP-H network is more suitable to be employed in the practical scenario since it can approach the optimal performance and reduce the computation complexity simultaneously.

%
%

\begin{figure*}[t]
  \centering
  \subfigure[The LAMP network with $M=1$]
  {\includegraphics[width=.43\textwidth]{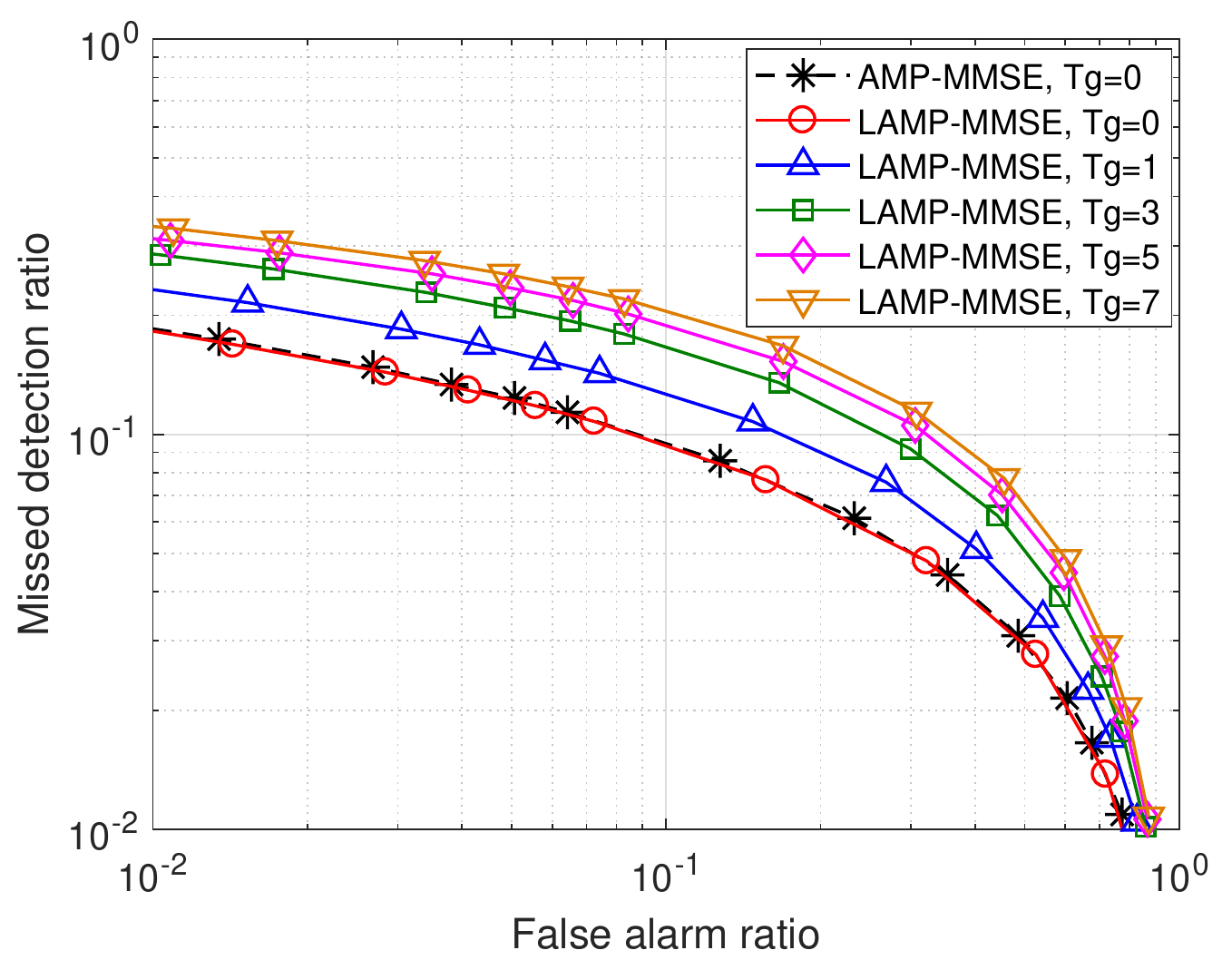}\label{Fig:LAMP-SMV_Tg}}
  \subfigure[The LAMP-D network with $M=4$]
  {\includegraphics[width=.43\textwidth]{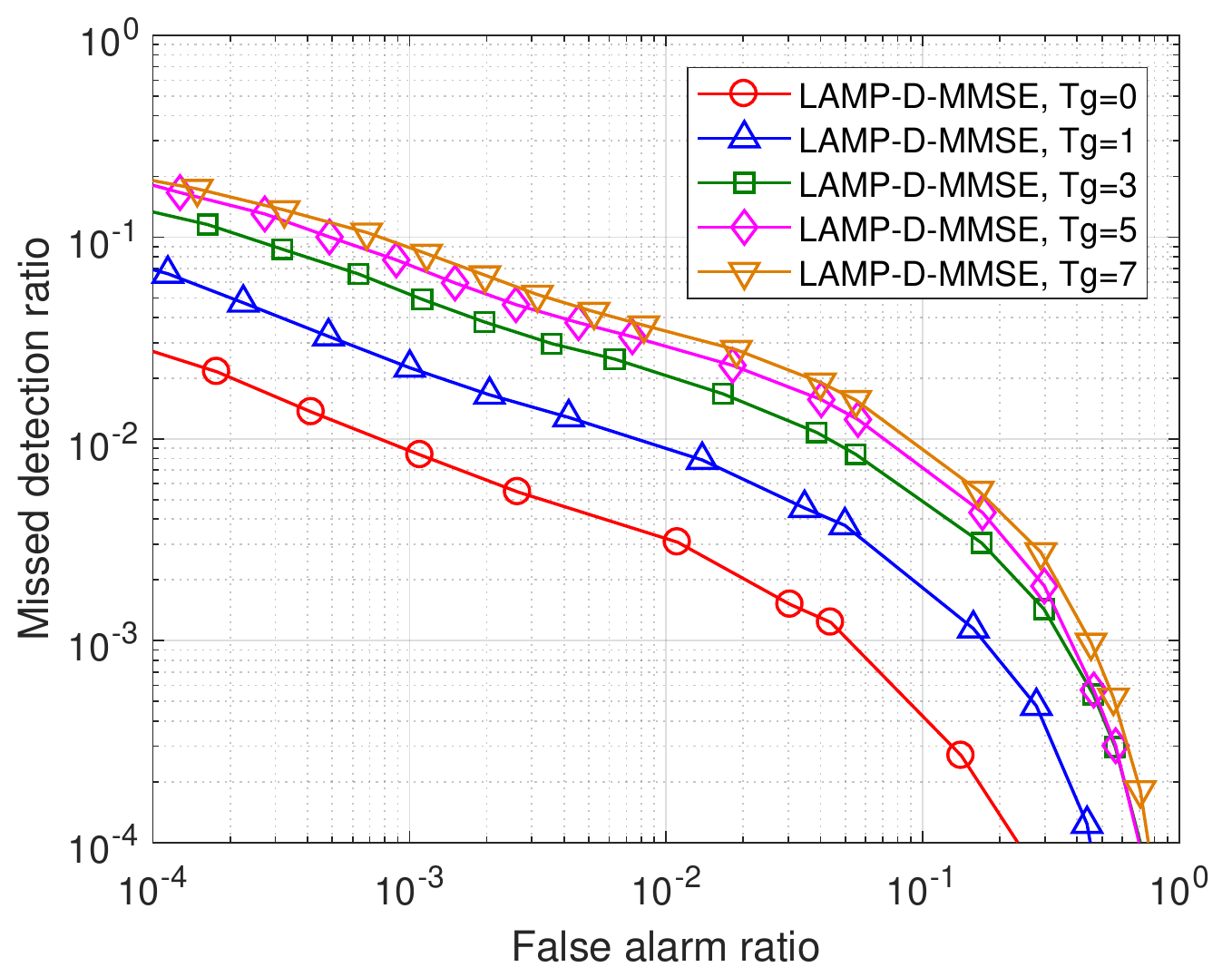}\label{Fig:LAMP-D_Tg}}
  \subfigure[The LAMP-C network with $M=4$]
  {\includegraphics[width=.43\textwidth]{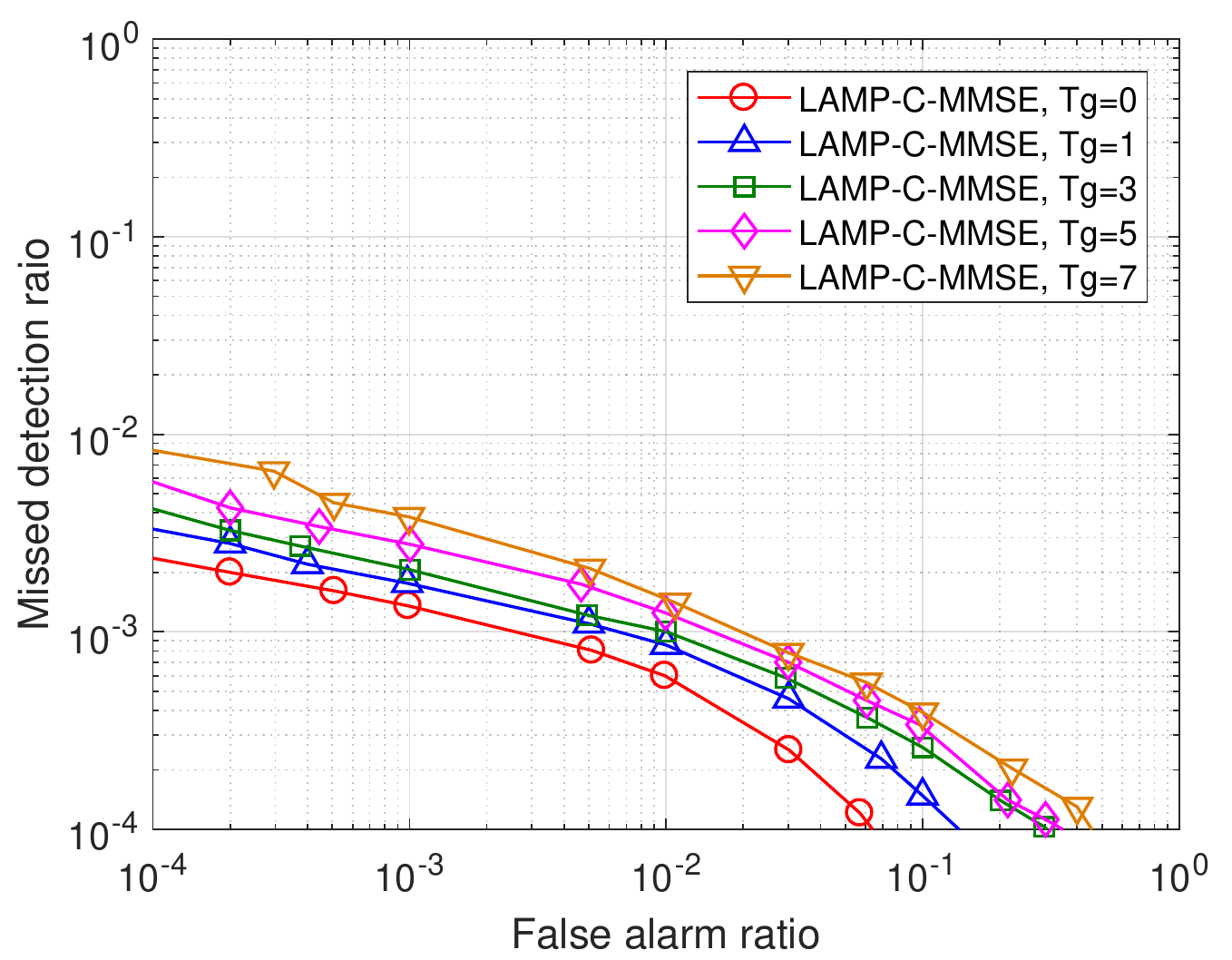}\label{Fig:LAMP-C_Tg}}
  \subfigure[The LAMP-H network with $M=4$]
  {\includegraphics[width=.43\textwidth]{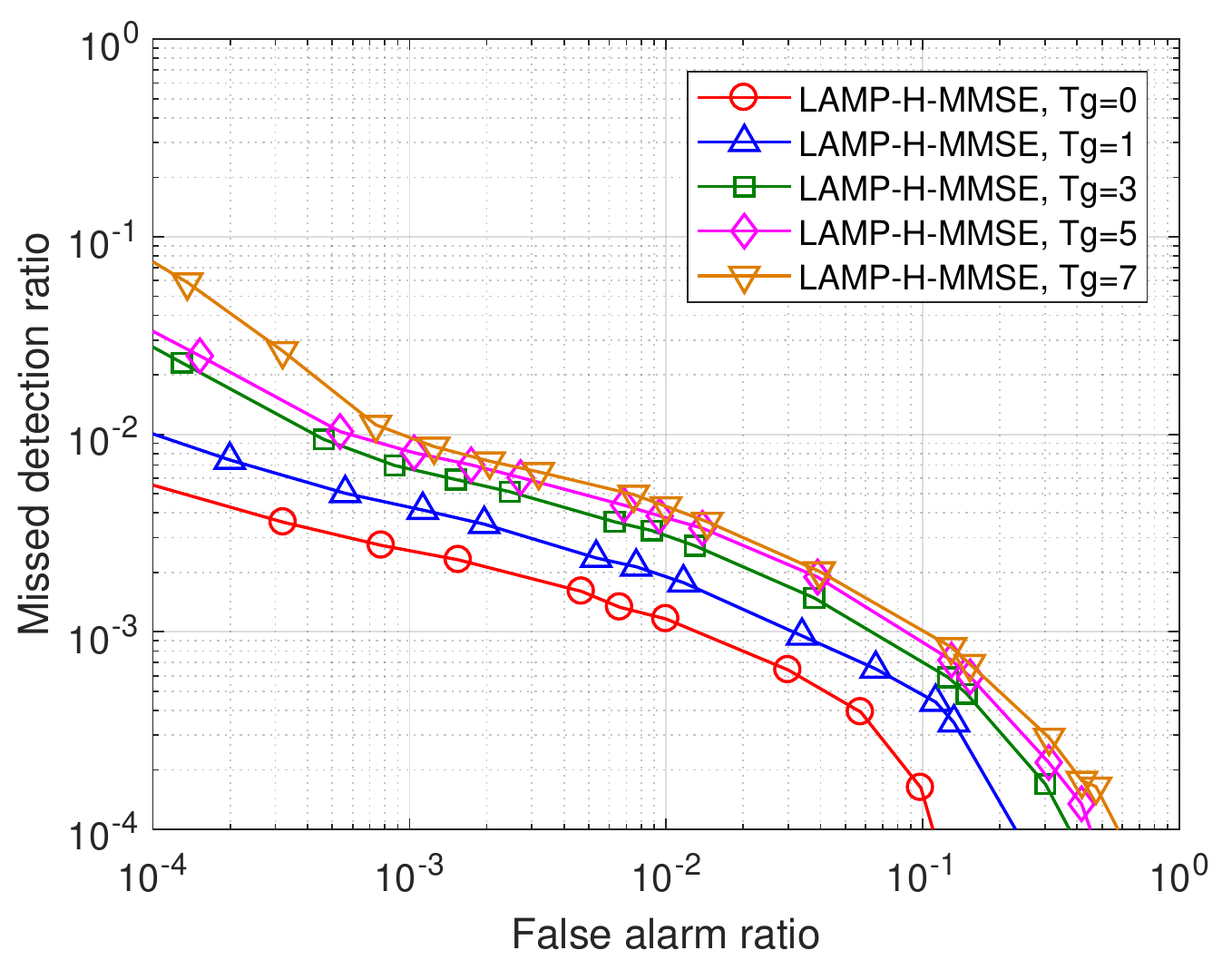}\label{Fig:LAMP-H_Tg}}
  \vspace{-0.4cm}
  \caption{The impact of the maximal symbol delay on the performance of the proposed LAMP networks in both the single-antenna scenario and multiple-antenna scenario.}\label{Fig:LAMP_Tg}
  \vspace{-1cm}
\end{figure*}


Finally, Fig. \ref{Fig:LAMP_Tg} evaluates the impact of the maximal delay spread on the user activity detection performance of the LAMP network with MMSE-optimal shrinkage function. For comparison purpose, the synchronous system is also included, which is a special case of the asynchronous system (i.e., $T_g=D=0$). Therefore, its performance serves as a lower bound. Both the single-antenna and multiple-antenna scenarios are considered. In the single-antenna scenario, the performance of the AMP algorithm with known statistical information is also evaluated. We can find that LAMP has similar performance to AMP when $T_g=0$, and the performance of the LAMP network is only slightly worse when $T_g$ increases, which indicates that the performance of the LAMP network is robust to the maximal symbol delay.
In the multiple-antenna scenario, the proposed networks also perform insensitively to the maximal symbol delay as well. In particular, the gap between two systems with different maximal symbol delays are much smaller in LAMP-C and LAMP-H compared with LAMP-D. Thus, the LAMP network with vector shrinkage function is more robust to the maximal symbol delay, which leads to less performance loss for the asynchronous massive access system.
Additionally, when the accurate maximal symbol delay is unknown in advance, we can set $T_g$ to be a larger value in the signal model without causing much performance degradation.

\vspace{-0.3cm}
\section{Conclusion}\label{sec:conclusion}

This work shows that combining deep learning techniques with compressed sensing is effective for asynchronous grant-free massive connectivity with the signal model inserted into a guard time.
Specifically, we propose to design the neural networks based on the AMP framework to jointly detect active users, detect their delays and estimate their channels with unknown system statistics of the channels and the users. Both the scenarios where the BS has single antenna and multiple antennas are considered.
We first design the LAMP network for the single-antenna scenario to exploit the potential of AMP by learning the parameters from the training data, and the neural network can slightly outperform the AMP algorithm with perfectly known system statistics. Furthermore, three LAMP network structures are proposed for the multiple-antenna scenario. In specific, the LAMP-D network can take advantage of distributed computation units to save the running time consumption, the LAMP-C network can improve the recoverability by exploit the common sparsity in the channel matrix and allow feasible performance analysis, and the LAMP-H network has balanced complexity and recoverability, which is suitable to more complicated systems. Simulation results show that the significant performance improvement is achieved by the proposed LAMP networks when perfect system statistics are unavailable. Additionally, the performance of proposed LAMP networks is also shown to be robust to the maximal delay spread of the asynchronous users.


\bibliographystyle{IEEEtran}
\bibliography{ref}

\begin{thebibliography}{10}
\providecommand{\url}[1]{#1}
\csname url@samestyle\endcsname
\providecommand{\newblock}{\relax}
\providecommand{\bibinfo}[2]{#2}
\providecommand{\BIBentrySTDinterwordspacing}{\spaceskip=0pt\relax}
\providecommand{\BIBentryALTinterwordstretchfactor}{4}
\providecommand{\BIBentryALTinterwordspacing}{\spaceskip=\fontdimen2\font plus
\BIBentryALTinterwordstretchfactor\fontdimen3\font minus
  \fontdimen4\font\relax}
\providecommand{\BIBforeignlanguage}[2]{{%
\expandafter\ifx\csname l@#1\endcsname\relax
\typeout{** WARNING: IEEEtran.bst: No hyphenation pattern has been}%
\typeout{** loaded for the language `#1'. Using the pattern for}%
\typeout{** the default language instead.}%
\else
\language=\csname l@#1\endcsname
\fi
#2}}
\providecommand{\BIBdecl}{\relax}
\BIBdecl

\bibitem{Zhu_2020_ICC}
W.~{Zhu}, M.~{Tao}, X.~{Yuan}, and Y.~{Guan}, ``Asynchronous massive
  connectivity with deep-learned approximate message passing,'' in \emph{Proc.
  IEEE Int. Conf. Commun.}, June 2020, pp. 1--6.

\bibitem{Andrew_2014_JSAC}
J.~G. {Andrews}, S.~{Buzzi}, W.~{Choi}, S.~V. {Hanly}, A.~{Lozano}, A.~C.~K.
  {Soong}, and J.~C. {Zhang}, ``What will 5{G} be?'' \emph{IEEE J. Sel. Areas
  Commun.}, vol.~32, no.~6, pp. 1065--1082, June 2014.

\bibitem{Bock_2018_Access}
C.~{Bockelmann} and \emph{et~al}, ``Towards massive connectivity support for
  scalable m{MTC} communications in 5{G} networks,'' \emph{IEEE Access},
  vol.~6, pp. 28\,969--28\,992, 2018.

\bibitem{Bock_2016_CM}
C.~{Bockelmann}, N.~{Pratas}, H.~{Nikopour}, K.~{Au}, T.~{Svensson},
  C.~{Stefanovic}, P.~{Popovski}, and A.~{Dekorsy}, ``Massive machine-type
  communications in 5{G}: physical and {MAC}-layer solutions,'' \emph{IEEE
  Commun. Mag.}, vol.~54, no.~9, pp. 59--65, Sep. 2016.

\bibitem{Liu_2018_SPM}
L.~{Liu}, E.~G. {Larsson}, W.~{Yu}, P.~{Popovski}, C.~{Stefanovic}, and E.~{de
  Carvalho}, ``Sparse signal processing for grant-free massive connectivity: A
  future paradigm for random access protocols in the internet of things,''
  \emph{IEEE Signal Process. Mag.}, vol.~35, no.~5, pp. 88--99, Sep. 2018.

\bibitem{Schepker_2013_ISWCS}
H.~F. {Schepker}, C.~{Bockelmann}, and A.~{Dekorsy}, ``Exploiting sparsity in
  channel and data estimation for sporadic multi-user communication,'' in
  \emph{Proc. Int. Symp. Wireless Commun. Syst.}, Aug 2013, pp. 1--5.

\bibitem{Wunder_2015_GCW}
G.~{Wunder}, P.~{Jung}, and M.~{Ramadan}, ``Compressive random access using a
  common overloaded control channel,'' in \emph{Proc. IEEE Global Commun. Conf.
  Workshops}, Dec 2015, pp. 1--6.

\bibitem{Chen_2018_TSP}
Z.~{Chen}, F.~{Sohrabi}, and W.~{Yu}, ``Sparse activity detection for massive
  connectivity,'' \emph{IEEE Trans. Signal Process.}, vol.~66, no.~7, pp.
  1890--1904, April 2018.

\bibitem{Liu_2018_TSP}
L.~{Liu} and W.~{Yu}, ``Massive connectivity with massive {MIMO}-{P}art {I}:
  Device activity detection and channel estimation,'' \emph{IEEE Trans. Signal
  Process.}, vol.~66, no.~11, pp. 2933--2946, June 2018.

\bibitem{Senel_2018_TCOM}
K.~{Senel} and E.~G. {Larsson}, ``Grant-free massive {MTC}-enabled massive
  {MIMO}: A compressive sensing approach,'' \emph{IEEE Trans. Commun.},
  vol.~66, no.~12, pp. 6164--6175, Dec 2018.

\bibitem{Sun_2019_TCOM}
Z.~{Sun}, Z.~{Wei}, L.~{Yang}, J.~{Yuan}, X.~{Cheng}, and L.~{Wan},
  ``Exploiting transmission control for joint user identification and channel
  estimation in massive connectivity,'' \emph{IEEE Trans. Commun.}, vol.~67,
  no.~9, pp. 6311--6326, Sep. 2019.

\bibitem{Fengler_2020_arxiv}
\BIBentryALTinterwordspacing
A.~Fengler, S.~Haghighatshoar, P.~Jung, and G.~Caire, ``Non-bayesian activity
  detection, large-scale fading coefficient estimation, and unsourced random
  access with a massive {MIMO} receiver,'' 2020. [Online]. Available:
  \url{https://arxiv.org/abs/1910.11266v2}
\BIBentrySTDinterwordspacing

\bibitem{Chen_2019_TWC}
Z.~{Chen}, F.~{Sohrabi}, and W.~{Yu}, ``Multi-cell sparse activity detection
  for massive random access: Massive {MIMO} versus cooperative {MIMO},''
  \emph{IEEE Trans. on Wireless Commun.}, vol.~18, no.~8, pp. 4060--4074, 2019.

\bibitem{Ke_2020_JSAC}
M.~{Ke}, Z.~{Gao}, Y.~{Wu}, X.~{Gao}, and K.~{Wong}, ``Massive access in
  cell-free massive mimo-based internet of things: Cloud computing and edge
  computing paradigms,'' \emph{IEEE J. Sel. Areas Commun.}, pp. 1--1, 2020.

\bibitem{Borg_2017_TSP}
M.~{Borgerding}, P.~{Schniter}, and S.~{Rangan}, ``{AMP}-inspired deep networks
  for sparse linear inverse problems,'' \emph{IEEE Trans. Signal Process.},
  vol.~65, no.~16, pp. 4293--4308, Aug 2017.

\bibitem{Donoho_2009_PNAS}
D.~L. Donoho, A.~Maleki, and A.~Montanari, ``Message-passing algorithms for
  compressed sensing,'' \emph{Proc. Nat. Acad. Sci.}, vol. 106, no.~45, pp.
  18\,914--18\,919, 2009.

\bibitem{Poly_2017_ISIT}
Y.~{Polyanskiy}, ``A perspective on massive random-access,'' in \emph{Proc IEEE
  Int. Symp. Inf. Theory}, June 2017, pp. 2523--2527.

\bibitem{Amalladinne_2020_arxiv}
\BIBentryALTinterwordspacing
V.~K. Amalladinne, A.~K. Pradhan, C.~Rush, J.-F. Chamberland, and K.~R.
  Narayanan, ``Unsourced random access with coded compressed sensing:
  Integrating {AMP} and belief propagation,'' 2020. [Online]. Available:
  \url{https://arxiv.org/abs/2010.04364}
\BIBentrySTDinterwordspacing

\bibitem{Shyianov_2020_JSAC}
V.~{Shyianov}, F.~{Bellili}, A.~{Mezghani}, and E.~{Hossain}, ``Massive
  unsourced random access based on uncoupled compressive sensing: Another
  blessing of massive mimo,'' \emph{IEEE J. Sel. Areas Commun.}, pp. 1--1,
  2020.

\bibitem{Sad_2020_TWC}
E.~{Sadeghabadi}, S.~M. {Azimi-Abarghouyi}, B.~{Makki}, M.~{Nasiri-Kenari}, and
  T.~{Svensson}, ``Asynchronous downlink massive mimo networks: A stochastic
  geometry approach,'' \emph{IEEE Trans. Wireless Commun.}, vol.~19, no.~1, pp.
  579--594, 2020.

\bibitem{Abebe_2017_ICC}
A.~T. {Abebe} and C.~G. {Kang}, ``Comprehensive grant-free random access for
  massive low latency communication,'' in \emph{Proc. IEEE Int. Conf. Commun.},
  May 2017, pp. 1--6.

\bibitem{Ama_2019_ICASSP}
V.~K. {Amalladinne}, K.~R. {Narayanan}, J.~{Chamberland}, and D.~{Guo},
  ``Asynchronous neighbor discovery using coupled compressive sensing,'' in
  \emph{Proc. IEEE Int. Conf. Acoustic, Speech, Signal Process.}, May 2019, pp.
  4569--4573.

\bibitem{Ding_2019_TWC}
T.~{Ding}, X.~{Yuan}, and S.~C. {Liew}, ``Sparsity learning-based multiuser
  detection in grant-free massive-device multiple access,'' \emph{IEEE Trans.
  Wireless Commun.}, vol.~18, no.~7, pp. 3569--3582, July 2019.

\bibitem{Bai_2019_VTC}
Y.~{Bai}, B.~{Ai}, and W.~{Chen}, ``Deep learning based fast multiuser
  detection for massive machine-type communication,'' in \emph{Proc. IEEE 90th
  Veh. Technol. Conf.}, Sep. 2019, pp. 1--5.

\bibitem{Zhang_2019_TVT}
Z.~{Zhang}, Y.~{Li}, C.~{Huang}, Q.~{Guo}, C.~{Yuen}, and Y.~L. {Guan},
  ``{DNN}-aided block sparse bayesian learning for user activity detection and
  channel estimation in grant-free non-orthogonal random access,'' \emph{IEEE
  Trans. Veh. Technol.}, vol.~68, no.~12, pp. 12\,000--12\,012, Dec 2019.

\bibitem{Schepker_2013_VTC}
H.~F. {Schepker}, C.~{Bockelmann}, and A.~{Dekorsy}, ``Coping with {CDMA}
  asynchronicity in compressive sensing multi-user detection,'' in \emph{Proc.
  IEEE 77th Veh. Technol. Conf.}, 2013, pp. 1--5.

\bibitem{Gregor_2010_ICML}
K.~Gregor and Y.~LeCun, ``Learning fast approximations of sparse coding,'' in
  \emph{Proc. Int. Conf. Mach. Learn.}, 2010, pp. 399--406.

\bibitem{Chen_2018_NIPS}
X.~Chen, J.~Liu, Z.~Wang, and W.~Yin, ``Theoretical linear convergence of
  unfolded ista and its practical weights and thresholds,'' in \emph{Proc.
  Neural Inf. Process. Syst. Conf.}, 2018, pp. 9061--9071.

\bibitem{Tropp_2006_SP}
J.~A. Tropp, ``Algorithms for simultaneous sparse approximation. part {II}:
  Convex relaxation,'' \emph{Signal Processing}, vol.~86, no.~3, pp. 589 --
  602, 2006.

\bibitem{Ziniel_2013_TSP}
J.~{Ziniel} and P.~{Schniter}, ``Efficient high-dimensional inference in the
  multiple measurement vector problem,'' \emph{IEEE Trans. Signal Process.},
  vol.~61, no.~2, pp. 340--354, Jan 2013.

\bibitem{Kim_2011_arxiv}
\BIBentryALTinterwordspacing
J.~Kim, W.~Chang, B.~C. Jung, D.~Baron, and J.~C. Ye, ``Belief propagation for
  joint sparse recovery,'' 2011. [Online]. Available:
  \url{http://arxiv.org/abs/1102.3289}
\BIBentrySTDinterwordspacing

\bibitem{Rume_1986_PDP}
D.~{Rumelhart}, G.~{Hinton}, and R.~{Williams}, ``Learning representations by
  back-propagating errors,'' \emph{Parallel Distributed Processing:
  Explorations in the Microstructure of Cognition}, vol.~1, pp. 318--362, 1986.

\bibitem{Metzler_2017_NIPS}
C.~Metzler, A.~Mousavi, and R.~Baraniuk, ``Learned {D-AMP}: Principled neural
  network based compressive image recovery,'' in \emph{Proc. Neural Inf.
  Process. Syst. Conf.}, 2017, pp. 1772--1783.

\bibitem{He_2019_WCM}
H.~{He}, S.~{Jin}, C.~{Wen}, F.~{Gao}, G.~Y. {Li}, and Z.~{Xu}, ``Model-driven
  deep learning for physical layer communications,'' \emph{IEEE Wireless
  Commun.}, vol.~26, no.~5, pp. 77--83, Oct. 2019.

\bibitem{Metzler_2016_TIT}
C.~A. {Metzler}, A.~{Maleki}, and R.~G. {Baraniuk}, ``From denoising to
  compressed sensing,'' \emph{IEEE Trans. Inf. Theory}, vol.~62, no.~9, pp.
  5117--5144, Sep. 2016.

\end{thebibliography}

\end{document}